\documentclass[compsoc,conference,a4paper,10pt,times]{IEEEtran}
\IEEEoverridecommandlockouts

\pdfoutput=1

\usepackage{fancyhdr}
\usepackage[normalem]{ulem}
\usepackage[hyphens]{url}
\usepackage{amsmath,amssymb,amsfonts}
\usepackage{multirow}
\usepackage{caption}
\usepackage{subcaption}
\usepackage{float}
\usepackage[hyphens]{url}
\usepackage{mathtools}
\usepackage{xspace}
\usepackage{color}
\usepackage{comment}
\usepackage[flushleft]{threeparttable}
\usepackage[table]{xcolor}
\usepackage[ruled,noend,vlined,linesnumbered]{algorithm2e}
\usepackage[noend]{algcompatible}
\usepackage{cite}
\usepackage{epstopdf}
\usepackage{pifont}
\usepackage{listings}
\usepackage{tikz}
\usepackage{enumitem}
\usepackage{rotating}
\usepackage{mdframed}
\usepackage{wasysym}
\usepackage{hyperref}
\usepackage[numbers,square,sort&compress]{natbib}

\def\BibTeX{{\rm B\kern-.05em{\sc i\kern-.025em b}\kern-.08em
    T\kern-.1667em\lower.7ex\hbox{E}\kern-.125emX}}

\SetAlCapNameFnt{\scriptsize}
\SetAlCapFnt{\scriptsize}

\newcommand{\bheading}[1]{{\vspace{2pt}\noindent{\textbf{#1}}\hspace{2pt}}}

\newcommand{\RNum}[1]{\uppercase\expandafter{\romannumeral #1\relax}}

\newcommand{\secref}[1]{\mbox{Sec.~\ref{#1}}\xspace}

\def\authnotes{1}
\newcommand{\authnote}[2]{\ifnum\authnotes=1\begin{quote}\textbf{#1 says:} #2\end{quote}\fi}
\newcommand{\fixme}[1]{\ifnum\authnotes=1\textbf{\textcolor{red}{[FIXME: #1]}}\fi}

\makeatletter
\newcommand{\removelatexerror}{\let\@latex@error\@gobble}
\makeatother

\newenvironment{packeditemize}{
\begin{list}{$\bullet$}{
\setlength{\labelwidth}{8pt}
\setlength{\itemsep}{0pt}
\setlength{\leftmargin}{\labelwidth}
\addtolength{\leftmargin}{\labelsep}
\setlength{\parindent}{0pt}
\setlength{\listparindent}{\parindent}
\setlength{\parsep}{0pt}
\setlength{\topsep}{3pt}}}{\end{list}}

\makeatletter
\def\@copyrightspace{\relax}
\makeatother

\renewcommand{\figurename}{Fig.}

\begin{document}
\title{Revisiting and Evaluating Software Side-channel Vulnerabilities and Countermeasures in Cryptographic Applications}

\author{\IEEEauthorblockN{Tianwei Zhang}
\IEEEauthorblockA{\textit{Nanyang Technological University} \\
tianwei.zhang@ntu.edu.sg}
\and
\IEEEauthorblockN{Jun Jiang}
\IEEEauthorblockA{\textit{Two Sigma Investments, LP} \\
jiangcj@pathsec.org}
\and
\IEEEauthorblockN{Yinqian Zhang}
\IEEEauthorblockA{\textit{The Ohio State University} \\
yinqian@cse.ohio-state.edu}
}

\maketitle

\definecolor{dkgreen}{rgb}{0,0.6,0}
\definecolor{gray}{rgb}{0.5,0.5,0.5}
\definecolor{mauve}{rgb}{0.58,0,0.82}

\lstset{frame=tb,
  language=C,
  aboveskip=3mm,
  belowskip=3mm,
  showstringspaces=false,
  columns=flexible,
  basicstyle={\small\ttfamily},
  numbers=left,
  numbersep=-2pt, 
  numberstyle=\tiny\color{gray},
  keywordstyle=\color{blue},
  commentstyle=\color{dkgreen},
  stringstyle=\color{mauve},
  breaklines=true,
  breakatwhitespace=true,
  tabsize=3
}

\begin{abstract}
We systematize software side-channel attacks with a focus on vulnerabilities
and countermeasures in the cryptographic implementations. Particularly, we survey
past research literature to categorize vulnerable implementations, and identify
common strategies to eliminate them. We then evaluate popular libraries and
applications, quantitatively measuring and comparing the vulnerability severity,
response time and coverage. Based on these characterizations and evaluations, we
offer some insights for side-channel researchers, cryptographic software developers
and users. We hope our study can inspire the side-channel research community to discover
new vulnerabilities, and more importantly, to fortify applications against them.
\end{abstract}

\section{Introduction}
\label{sec:intro}




Side-channel attacks have become a severe threat to computer applications and systems.
They exploit the vulnerabilities in the implementations instead of the algorithms.
Vulnerable implementations can exhibit input-dependent non-functional behaviors at
runtime, which can be observed by an adversary to fully or partially recover the
sensitive input. Over the past few years, numerous side-channel vulnerabilities were
discovered and exploited to defeat modern cryptographic schemes, allowing adversaries to
break strong ciphers in a short period of time with very few trials.

Defeating side-channel vulnerabilities has been a long-standing goal for providing robust
cryptographic protection. Although security-aware systems \cite{ShSoCh:11, KiPeMa:12, VaDaSh:11,
ZhRe:13, LiGaRe:14, ZhZhLe:16} and architectures \cite{WaLe:07, WaLe:08, DoJaLo:12, LiLe:14,
LiGeYa:16} were designed to mitigate side-channel attacks, a more efficient and practical approach
is to eliminate side-channel sources from software implementations.
Various tools and methods were proposed to facilitate creating software free of side channels
\cite{MoPiSc:05, CoVeDe:09} or verifying their non-existence \cite{BaBeGu:14, AlBaBa:16, BaPiTr:17, DeFrHo:17,
ReBaVe:17, BoHaKa:17}. It is however still very challenging to remove all side-channel
vulnerabilities from critical implementations, since cryptographic applications usually have a
large code base and high performance requirement. As such, the arms race between side-channel
attacks and defenses remains heated.

Past several decades have seen a large amount of literature about side-channel
vulnerabilities. Meanwhile, various open-source libraries and commercial products have
introduced different mitigation solutions. Thus, it becomes necessary to systematize
the knowledge about the characteristics and evolution of these vulnerabilities and countermeasures.
We are particularly interested in three questions: (1) \emph{What are the common and distinct features
of various vulnerabilities?} (2) \emph{What are common mitigation strategies?} (3) \emph{What is the
status quo of cryptographic applications regarding side-channel vulnerabilities?} Past work
only surveyed attack techniques and media \cite{ZaArBr:07, XuSoJi:16, Na:16, HuAlZa:16,
BiGhNa:17, Tu:17, UlZsFa:17, BeWeMu:17, GeYaCo:18, Sz:18, SpMoKo:18, Mo:18}, without
offering unified summaries for software vulnerabilities and countermeasures that are more
useful.

This paper provides a comprehensive characterization of side-channel vulnerabilities and 
countermeasures, as well as evaluations of cryptographic applications related to
side-channel attacks. We present this study in three directions. (1) Systematization of 
literature: we characterize the vulnerabilities from past work with regard to the 
implementations; for each vulnerability, we describe the root cause and the technique
required to launch a successful attack. (2) Identification and abstraction of key countermeasure 
techniques: we summarize the common strategies to mitigate different categories of vulnerabilities;
we also explore and compare the effectiveness of each countermeasure implementation under different 
threat models. (3) Evaluation of cryptographic applications: we perform a
timeline analysis of side-channel vulnerabilities and the corresponding patches in various libraries
and products, and evaluate the vulnerability severity, patch release speed and coverage from a practical
perspective.

\bheading{Scope.}
There are generally two types of side-channel attacks.
In \emph{software attacks}, an adversary interacts with the victim application through a local malicious program
or over the network to collect information such as execution time \cite{Be:05, BoMi:06}
and memory access pattern \cite{OsShTr:06} for recovering the victim's secrets. Such vulnerabilities 
are usually caused by critical control flow or data flow leakage. In \emph{physical attacks}, an 
adversary physically interferes with the victim's execution (e.g., fault injection
\cite{BaBrKo:12}) or approaches the victim to collect physical signals such as 
acoustic emission \cite{GeShTr:14}, electromagnetic radiation \cite{GePiTr:15, GePaPi:15} 
and power trace \cite{Co:99, ArTh:07}. The adversary utilizes special analysis (e.g.,
power analysis \cite{Ma:02, KoJaJu:98}) to obtain finer grained information
(e.g., intermediate values, Hamming weights) than control flow or data flow. In this paper, we 
mainly focus on software attacks, which are more exploitable and common. Physical attacks and 
vulnerabilities are out of the scope of this paper.

\bheading{Contributions.}
The main purpose of this work is to help researchers, software developers and users better 
understand the status quo and future direction of side-channel research and 
countermeasure development. Based on our systematization, we: 1) propose three possible 
directions for researchers to consider in their future exploration; 2) provide three
recommendations for developers to follow in security enhancement of their 
applications against side-channel attacks; 3) identify three indications for 
users to utilize in selecting libraries and implementations
most suitable for their usage scenarios. The key contributions of this paper are:

\begin{packeditemize}
\item Characterization of side-channel vulnerabilities in implementations of cryptographic operations (\secref{sec:crypto}).
\item Identification and dissection of common countermeasure technique designs (\secref{sec:insights}).
\item Evaluation of cryptographic applications, and analysis of vulnerabilities and countermeasures (\secref{sec:history}). 
\item Insights and recommendations for side-channel researchers, software developers and users (\secref{sec:lesson}).
\end{packeditemize}


\section{Background}
\label{sec:bg}

\subsection{Basics of Cryptography}

\subsubsection{Asymmetric Cryptography} Each user has a public key that is widely distributed
and a private key that is kept to herself. This pair of keys can be used for data encryption/decryption
and digital signature.

\bheading{RSA \cite{RiShAd:78}.}
Two different large prime numbers, $p$ and $q$, and an integer $e$, are chosen to satisfy that
$gcd(p-1, e)=1$ and $gcd(q-1, e)=1$. Let $N = pq$ and $d = e^{-1} \mod (p-1)(q-1)$, then the
public key is the tuple $(N, e)$, and the private key is $d$. For a message $m$,
the ciphertext is calculated as $c \equiv m^e \mod N$.
For a ciphertext $c$, the message is decrypted as $m \equiv c^d \mod N$.
The security of RSA relies on the difficulty of factoring large integers.

\bheading{ElGamal \cite{El:85}.}
A large cyclic group $G$ is chosen. Let $q$ be its order, $g$ be its generator, $x$ be a random
positive integer smaller than $q$, and $h = g^x$, then the public key is the tuple $(G, q, g, h)$,
and the private key is $x$. To encrypt a message $m$, one chooses a random positive integer smaller
than $q$, denoted by $y$, and calculate the ciphertext as $(g^y, m \cdot h^y)$. To decrypt a
ciphertext $(c_1, c_2)$, one can compute $m = c_2 \cdot (c_1^x)^{-1}$. The security of ElGamal
relies on the difficulty of solving the Discrete Logarithm Problem.

\bheading{Elliptic Curve Cryptography (ECC) \cite{Mi:85, Ko:87}.}
An elliptic curve group is chosen with prime order $p$ and generator $G$. Let $k$ be a random
positive integer smaller than $p$, then $k$ is the private key, and the public key is $D=kG$. To
encrypt a message $m$, one encodes $m$ to a point $M$ on the curve, chooses a random positive integer
$r$, and calculate the ciphertext as two points $(M + rD, rG)$. To decrypt a ciphertext $(C_1, C_2)$,
one can compute $M= C_1 – kC_2$, and decode point $M$ to message $m$. The security of ECC relies on the
difficulty of solving the Elliptic Curve Discrete Logarithm Problem. Appendix \ref{sec:append-ecc} has more details.



\subsubsection{Symmetric Cryptography}
A single key is used in both data encryption and decryption, and is
shared between two users. Digital signature cannot supported.

\bheading{AES-128.}
During key setup, a 16-byte secret key
$k = (k_0,...,k_{15})$ is expanded into 11 round keys while
the initial round key is just the original key itself.
Given a 16-byte plaintext $p=(p_0,...,p_{15})$, encryption proceeds by computing a 16-byte intermediate
state $x^{(r)}=(x_0^{(r)},...,x_{15}^{(r)})$ at each round $r$ as a $4 \times 4$ matrix.
The initial state is computed as $x_i^{(0)} = p_i\oplus k_i$ $(i=0,...,15)$, known as \texttt{AddRoundKey}.
Each of the following 9 rounds consists of \texttt{SubBytes} (byte substitution based on a lookup table),
\texttt{ShiftRows} (transposition of bytes within each of the three last rows in the $4 \times 4$ matrix),
\texttt{MixColumns} (matrix multiplication to make each byte represents a weighted sum of
all bytes in its column) and \texttt{AddRoundKey} operations.
The final round has only \texttt{SubBytes}, \texttt{ShiftRow} and \texttt{AddRoundKey}
operations, and its output is the ciphertext.


\subsubsection{Post-Quantum Cryptography}
This cipher family was proposed to sustain attacks by a quantum computer. One popular scheme is lattice-based
cryptography.

\bheading{NTRU \cite{HoPiSi:98}.}
It utilizes simple polynomial multiplication in the ring of truncated polynomials $\mathbb{Z}_q[X]/(X^N-1)$.
For encryption, the private key consists of a pair of polynomials $f$ and 
$g$, and the public key $h = p \cdot f_q^{-1} \cdot g \mod q$, where $f_q^{-1}$ denotes
the inverse of $f$ modulo $q$. To encrypt a message $m$, one needs to compute a hash function
$r = G(m)$ and $m'=m\oplus H(r \cdot h \mod q)$. Then the ciphertext is $e=(r \cdot h+m')\mod q$. To decrypt the message, one needs to 
first recover $m' = ((f \cdot e \mod q)\mod p) \cdot f_p^{-1}\mod p$ and then the plaintext is computed as $M=m'\oplus H(e-m'\mod q)$. 
The security of NTRU relies on the difficulty of solving the shortest vector problem in a lattice.


\bheading{Ring Learning With Errors (RLWE) \cite{LyPeRe:10}.}
To generate a key pair, one needs to create a polynomial $a\in
\mathbb{R_q}$ with coefficients chosen uniformly in $\mathbb{Z}_q$, sample two polynomials $r_1, r_2
\in \mathbb{R}_q$ from $\chi$ and compute $p=r_1-a\cdot r_2 \in \mathbb{R}_q$. Then the public key is $(a,p)$
and the private key is $r_2$. To encrypt a message $m$, one needs to first encode $m$ to a
polynomial $\overline{m}$, and sample three polynomials $e_1, e_2, e_3 \in \mathbb{R}_q$ from $\chi$.
Then the ciphertext is $(c_1, c_2)$, where $c_1 = a\cdot e_1 + e_2$ and $c_2= p\cdot e_1 + e_3 + \overline{m}$.
To decrypt the message, one needs to compute $m'=c_1 \cdot r_2 + c_2 \in \mathbb{R}_q$ and decode the coefficients of $m'$
to either 0 or 1.

\bheading{Bimodal Lattice Signature Scheme (BLISS) \cite{DuDuLe:13}.}
The private key is $S=(s_1,s_2)$ where $s_1,s_2 \in \mathbb{Z}_q[X]/(X^N-1)$
and the corresponding public key is $A=(2a_q, q-2) \mod 2q$ where $a_q = s_2/s_1$. To sign a
message $\mu$, two blinding values $y_1, y_2 \in \mathbb{R}$ are sampled from a discrete Gaussian
distribution. A hash value is computed as $c = H(\lfloor u \rceil_d \mod p, \mu)$, where $u=\zeta \cdot 2a_qy_1 + y_2 \mod 2q$,
and $\zeta = (1/(q-2)) \mod 2q$. The signature is then the triple $(c, z_1, z_2)$, with $z_i = y_i + (-1)^b s_i\cdot c \mod 2q$,
where $b$ is a random bit.
Signature verification is performed by checking $c=H(\lfloor \zeta \cdot 2a_q \cdot z_1 + \zeta \cdot q \cdot c\rceil_d + z_2 \mod p, \mu)$.

\subsubsection{Cryptographic Protocol}
\label{sec:protocol-pad}
SSL/TLS allows a server and a client to use the handshake protocol to exchange a symmetric key $K$,
which they use for later secure communications. 
More details about the protocol can be found in Appendix \ref{sec:append-pad}.

\bheading{Key exchange.}
RSA is commonly adopted to exchange the symmetric key $K$, following Public Key 
Cryptography Standards (PKCS). The client generates a random non-zero padding
string \texttt{pad} that is at least 8 bytes, creates a block $0x00||0x02||\texttt{pad}||0x00||K$,
encrypts it using RSA and sends the ciphertext to the server.
When the server decrypts the ciphertext, she accepts this message only when the first two
bytes are $0x00$ and $0x02$, and next 8 bytes are non-zero. Then she searches for $0x00$ in
the remaining data, and everything after that will be the key.

\bheading{Hash and encryption.}
After the key is established, standard network protocols adopt CBC-MAC
to encrypt messages. (1) Message Authentication Code (MAC) is calculated over the sequence
number \texttt{SEQ}, header \texttt{HDR} and message $m$. (2) The plaintext $P$ is created by
concatenating $m$, the MAC from step (1), and a padding string \texttt{pad}, which
is chosen to make the byte length of $P$ a multiple of the block size $b$. (3) $P$ is divided
into blocks of $b$ bytes, each block encrypted with key $K$. (4) The final message is the
concatenation of \texttt{HDR} and all encrypted blocks.
The receiver decrypts the ciphertext in CBC mode and validates the padding format and
the MAC. If both are correct, she accepts the original intact message $m$.


\subsection{Basics of Side-channel Attacks}
Side-channel attacks enable adversaries to steal secrets by exploiting observable information from
the application's execution. When the application takes a secret input
$\mathcal{X}$, the host system or application shows a runtime characteristic
$\mathcal{Y}$, which can be captured by an adversary. By identifying the correlation
$\mathcal{Y} \sim \mathcal{X}$, the adversary is able to infer the secret input via the
side-channel information.

\subsubsection{Vulnerabilities}
\begin{figure}[t]
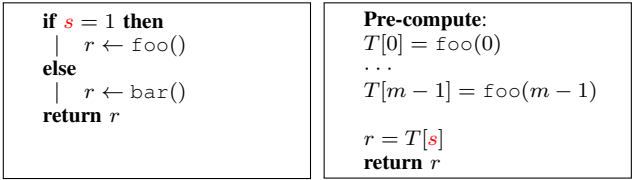

  \centering
  \removelatexerror
  \begin{minipage}[t][][b]{.49\linewidth}
    \begin{algorithm*}[H]
      \footnotesize
      \SetAlgoLined
      \Indp
      \Indm
      \SetKwProg{myalg}{function}{}{end}
        \eIf{$\textcolor{red}{s}=1$} {
          $r\gets \texttt{foo}()$
        }{
          $r\gets \texttt{bar}()$
        }
        \KwRet{$r$} \\
        $\;$
        $\;$ \\
        $\;$        
     {}
    \end{algorithm*}
  \captionsetup{labelformat=empty}
  \subcaption{\footnotesize{Control flow vulnerability}}
  \label{fig:insecure}
  \end{minipage} \hfill
  \begin{minipage}[t][][b]{.49\linewidth}
    \begin{algorithm*}[H]
      \footnotesize
      \SetAlgoLined
      \Indp
      \Indm
      \SetKwProg{myalg}{function}{}{end}
        \textbf{Pre-compute}: \\
        $T[0] = \texttt{foo}(0)$ \\
        $\cdot\cdot\cdot$ \\
        $T[m-1] = \texttt{foo}(m-1)$ \\
        $\;$ \\
        $r = T[\textcolor{red}{s}]$ \\
        \KwRet{$r$} \\
      {}
    \end{algorithm*}
  \captionsetup{labelformat=empty}
  \subcaption{\footnotesize{Data flow vulnerability}}
  \label{fig:memory-traffic}
  \end{minipage}
  \caption{Side-channel vulnerabilities}
  \label{fig:sc-vul}
  \vspace{-15pt}
\end{figure}

The root cause of side-channel attacks is that application's runtime behavior depends
on the secrets. Generally there are two types of leakage sources:

\bheading{Secret-dependent control flow.} When the secret $s$ is different, the program
executes a different code path and invokes a different function (Figure \ref{fig:insecure}).
This yields different runtime behaviors distinguishable by the adversary.

\bheading{Secret-dependent data flow.}
A program may access data whose locations are determined by the secret $s$
(Figure \ref{fig:memory-traffic}). The memory access pattern allows the adversary to
make educated guesses about the secret value.

\RestyleAlgo{boxed}
\SetAlgoLined
\LinesNumberedHidden

\subsubsection{Techniques}
Another key factor for a successful side-channel attack is the technique to capture 
useful characteristics of the application or the system. We consider two types of
software-based techniques and approaches.


\bheading{Network-level attacks.}
The remote adversary connects to the victim application via networks. Thus, side-channel
information may exist in the responses from the victim such as message content
\cite{Bl:98, Va:02}, packet size \cite{ChWaWa:10} and response time \cite{MeSoJu:14, AlPa:12}.


\bheading{Host-level attacks.}
The malicious program and the victim application run on the same platform, and thus share
the same micro-architectural units.
Contention on these units can disclose the victim's
runtime behaviors to the adversary. Researchers have designed a large quantity of attack
techniques based on various hardware units including CPU cache \cite{Pe:05, OsShTr:06}, branch
predictor \cite{AcGuSe:07} and TLB \cite{GrRaBo:18}.

\section{Characterization of Vulnerabilities}
\label{sec:crypto}

We systematically characterize side-channel vulnerabilities from past work based on 
different operations in different cryptographic ciphers and protocols. Table
\ref{table:summary_crypto} summarizes the vulnerabilities we will describe.
For each vulnerability, we present the vulnerable operations, causes and the corresponding
attacks (types, techniques, side-channel information granularities).



\begin{table*}[ht]
\centering
\caption{Side-channel vulnerabilities. ($\blacksquare$: control flow, $\square$: data flow; $\CIRCLE$: host-level, $\Circle$: network-level)}
\label{table:summary_crypto}
\resizebox{\linewidth}{!}{
\begin{threeparttable}
\begin{tabular}{|c|c|c|c|c|c|c|c|c|}
  \hline
  \textbf{Category} & \textbf{Operation} & \textbf{Implementation} & \textbf{Application} & \textbf{Cause} & \textbf{Type} & \textbf{Granularity} & \textbf{Ref} & \textbf{Index}\\
  \hline
  \hline

  & \multirow{1}{*}{Modular Multiplication} & \multirow{1}{*}{Basic and Karatsuba multiplication} & RSA & $\blacksquare$ & $\Circle$ & Execution time & \cite{BrBo:05} & 1.1.1.a\\
  \cline{2-9}


  & & \multirow{2}{*}{Square-and-Multiply} & RSA & $\blacksquare$ & $\CIRCLE$ & Cacheline (\textsc{Flush-Reload}) & \cite{YaFa:14} & 1.2.1.a \\
  \cline{4-9}
  & & \multirow{2}{*}{Double-and-Add} & ElGamal & $\blacksquare$ & $\CIRCLE$ & Cacheline (\textsc{Prime-Probe}) & \cite{ZhJuRe:12, LiYaGe:15} & 1.2.1.b\\
  \cline{4-9}
  & & & EdDSA & $\blacksquare$ & $\CIRCLE$ & Memory Page (TLB) & \cite{GrRaBo:18} & 1.2.1.c\\
  \cline{3-9}

  & & \multirow{1}{*}{Square-and-Multiply-always} & RSA & $\blacksquare$ & $\CIRCLE$ & Branch & \cite{DoKo:17} & 1.2.2.a\\
  \cline{4-9}
   & & \multirow{1}{*}{Double-and-Add-always} & RSA & $\blacksquare$ & $\CIRCLE$ & Memory Page (TLB) & \cite{GrRaBo:18} & 1.2.2.b\\
  \cline{3-9}
   & & \multirow{7}{*}{Sliding window} & RSA & $\blacksquare$ & $\CIRCLE$ & Cacheline (\textsc{Prime-Probe}) & \cite{Pe:05} & 1.2.3.a \\
  \cline{4-9}
   & & & RSA & $\blacksquare$ & $\CIRCLE$ & Cacheline (\textsc{Flush-Reload}) & \cite{BeBrGe:17} & 1.2.3.b\\
  \cline{4-9}
  & & & ECDSA & $\blacksquare$ & $\CIRCLE$ & Cacheline (\textsc{Flush-Reload}) & \cite{BeVaSm:14, VaSmYa:15, FaWaCh:16, AlBrFa:16} & 1.2.3.c\\
  \cline{4-9}
  & & & ECDSA & $\blacksquare$ & $\CIRCLE$ & micro-operation (Execution Port) & \cite{AlBrHa:19} & 1.2.3.d\\
  \cline{4-9}
  Asymmetric & Modular Exponentiation  & & RSA & $\square$ & $\CIRCLE$ & Cacheline (\textsc{Prime-Probe}) & \cite{InGuIr:16} & 1.2.3.e\\
  \cline{4-9}
  Cryptography & Scalar Multiplication & & ElGamal & $\square$ & $\CIRCLE$ & Cacheline (\textsc{Prime-Probe}) & \cite{LiYaGe:15} & 1.2.3.f\\
  \cline{4-9}
  & & & ECDSA & $\square$ & $\CIRCLE$ & Cacheline (\textsc{Prime-Probe}) & \cite{BrHa:09} & 1.2.3.g\\
  \cline{3-9}

  & & Fixed window & RSA & $\square$  & $\CIRCLE$ & Cache bank & \cite{YaGeHe:17} & 1.2.4.a\\
  \cline{3-9}

  & & \multirow{2}{*}{Montgomery ladder} & ECDSA & $\blacksquare$ & $\CIRCLE$ & Cacheline (\textsc{Flush-Reload}) & \cite{YaBe:14} & 1.2.6.a\\
  \cline{4-9}
  & &  & ECDSA & $\blacksquare$ & $\Circle$ & Execution time & \cite{BrTu:11} & 1.2.6.b\\
  \cline{3-9}

  & & \multirow{3}{*}{Branchless montgomery ladder}  & ECDH & $\square$ & $\CIRCLE$ & Cacheline (\textsc{Flush-Reload}) & \cite{ShKiKw:18} & 1.2.6.a\\
  \cline{4-9}
  & & & ECDH & $\blacksquare$ & $\CIRCLE$ & Cacheline (\textsc{Flush-Reload}) & \cite{GeVaYa:17} & 1.2.7.b \\
  \cline{4-9}
  & & & ECDH & $\blacksquare$ & $\Circle$ & Execution time & \cite{KaPeVa:16} & 1.2.7.c\\
  \cline{2-9}

  & \multirow{3}{*}{Modular Inverse} & \multirow{3}{*}{Binary Extended Euclidean Algorithm} & RSA & $\blacksquare$ & $\CIRCLE$ & Branch & \cite{AcGuSe:07} & 1.3.1.a\\
  \cline{4-9}
  & & & RSA & $\blacksquare$ & $\CIRCLE$ & Memory Page (Controlled-channel)& \cite{WeSpBo:18} & 1.3.1.b\\
  \cline{4-9}
  & & & RSA & $\blacksquare$ & $\CIRCLE$ & Cacheline (\textsc{Flush-Reload}) & \cite{AlGaTa:18} & 1.3.1.c\\
  \hline

  & \multirow{4}{*}{Substitution-Permutation}& \multirow{4}{*}{T-box lookup} & AES & $\square$ & $\CIRCLE$ & Cacheline (\textsc{Prime-Probe}) & \cite{OsShTr:06} & 2.1.1.a\\
  \cline{4-9}
  Symmetric & &  & AES & $\square$ &  $\CIRCLE$ & Cacheline (\textsc{Evict-Time}) & \cite{OsShTr:06} & 2.1.1.b\\
  \cline{4-9}
  Cryptography & &  & AES & $\square$ & $\CIRCLE$ & Cache (\textsc{Flush-Reload}) & \cite{GuBaKr:11, IrInEi:14} & 2.1.1.c\\
  \cline{4-9}
  & &  & AES & $\square$ & $\Circle$ & Execution time & \cite{Be:05, BoMi:06} & 2.1.1.d\\
  \hline

  \multirow{4}{*}{Post-Quantum}  & \multirow{4}{*}{Distribution Sampling}& \multirow{1}{*}{Cumulative Distribution Table sampling} & BLISS & $\square$ & $\CIRCLE$ & Cacheline (\textsc{Flush-Reload}) & \cite{BrHuLa:16,PeBrYa:17} & 3.1.1.a \\
  \cline{3-9}
  \multirow{4}{*}{Cryptography} & & \multirow{2}{*}{Rejection sampling} & BLISS & $\square$ & $\CIRCLE$ & Cacheline (\textsc{Flush-Reload}) & \cite{BrHuLa:16,PeBrYa:17} & 3.1.2.a\\
  \cline{4-9}
  & & & BLISS & $\blacksquare$ & $\CIRCLE$ & Branch & \cite{EsFoGr:17, TiWa:19} & 3.1.2.b \\
  \cline{2-9}

  & \multirow{1}{*}{Failure Rate Reduction} & Error Correcting Code & Ring-LWE & $\blacksquare$ & $\Circle$ & Execution time & \cite{DaTiVe:19} & 3.2.1.a \\
  \cline{2-9}
  & \multirow{1}{*}{Message Randomization} & Padding-Hash & NTRU & $\blacksquare$ & $\Circle$ & Execution time & \cite{SiWh:07} & 3.3.1.a\\
  \hline

  & \multirow{11}{*}{RSA-PAD} & \multirow{4}{*}{Error message} & SSL & $\blacksquare$ & $\Circle$ & Network message & \cite{Bl:98, BaFoKa:12} & 4.1.1.a\\
  \cline{4-9}
  & & & SSLv3.0, TLSv1.0 & $\blacksquare$ & $\Circle$ & Network message & \cite{KiPoFRo:03} & 4.1.1.b\\
  \cline{4-9}
  & & & TLS & $\blacksquare$ & $\Circle$ & Network message & \cite{HaJuCr:18} & 4.1.1.c\\
  \cline{4-9}
  & & & IKE & $\blacksquare$ & $\Circle$ & Network message & \cite{FeGrSc:18} & 4.1.1.d\\
  \cline{3-9}

  & & \multirow{7}{*}{Uniform response message} & TLSv1.0 & $\blacksquare$ & $\Circle$ & Execution time & \cite{MeSoJu:14} & 4.1.2.a\\
  \cline{4-9}
  & & & SSLv2.0 & $\blacksquare$ & $\Circle$ & Network message & \cite{AvScSo:16} & 4.1.2.b\\
  \cline{4-9}
  & & & TLS & $\blacksquare$ & $\CIRCLE$ & Page, Cacheline, Branch & \cite{XiLiCh:17} & 4.1.2.c\\
  \cline{4-9}
  & & & TLS & $\blacksquare$ & $\CIRCLE$ & Cacheline (\textsc{Flush-Reload}), Branch & \cite{RoGiGe:19} & 4.1.2.d\\
  \cline{4-9}
  & & & XML Encryption & $\blacksquare$ & $\Circle$ & Network message & \cite{JaScSo:12} & 4.1.2.e\\
  \cline{4-9}
  & & & XML Encryption & $\blacksquare$ & $\Circle$ & Execution time & \cite{JaScSo:12} & 4.1.2.f\\
  \cline{4-9}
  Cryptographic & & & XML Encryption & $\blacksquare$ & $\CIRCLE$ & Cacheline (\textsc{Flush-Reload}) & \cite{ZhJuRe:14} & 4.1.2.g\\
  \cline{2-9}

  Protocol & \multirow{11}{*}{CBC-MAC-PAD} & \multirow{5}{*}{Error message} & IPSec, WTLS, SSH2 & $\blacksquare$ &$\Circle$ & Network message & \cite{Va:02} & 4.2.1.a\\
  \cline{4-9}
  & & & SSLv3.0 & $\blacksquare$ & $\Circle$ & Network message & \cite{MoDuKo:14} & 4.2.1.b\\
  \cline{4-9}
  & & & IPSec & $\blacksquare$ & $\Circle$ & Network message  & \cite{DePa:07, DePa:10} & 4.2.1.c\\
  \cline{4-9}
  & & & Web apps & $\blacksquare$ & $\Circle$ & Network message & \cite{RiDu:10} & 4.2.1.d\\
  \cline{4-9}
  & & & ASP.NET & $\blacksquare$ & $\Circle$ & Network message  & \cite{DuRi:11} & 4.2.1.e\\
  \cline{3-9}

  & & \multirow{2}{*}{Uniform response message} & TLSv1.0 & $\blacksquare$ & $\Circle$ & Execution time & \cite{CaHiVa:03} & 4.2.2.a\\
  \cline{4-9}
  & & & TLSv1.1 & $\blacksquare$ & $\Circle$ & Execution time & \cite{AlPa:12} & 4.2.2.b\\
  \cline{3-9}
  & & Dummy MAC checking & TLSv1.0, TLSv1.1 & $\blacksquare$ & $\Circle$ & Execution time & \cite{FaPa:13} & 4.2.3.a\\
  \cline{3-9}

  & & \multirow{3}{*}{Constant-time compression} & TLSv1.1, TLSv1.2 & $\blacksquare$ & $\CIRCLE$ & Cacheline (\textsc{Flush-Reload}) & \cite{IrLnEi:15} & 4.2.4.a\\
  \cline{4-9}
  & &  & TLS & $\blacksquare$ & $\CIRCLE$ & Page, Cacheline, Branch & \cite{XiLiCh:17} & 4.2.4.b\\
  \cline{4-9}
  & &  & TLS & $\blacksquare$ & $\CIRCLE$ & Cacheline (\textsc{Prime-Probe}) & \cite{RoPaSh:18} & 4.2.4.c\\

  \hline
  \end{tabular}
  \end{threeparttable}}
  \vspace{-15pt}
\end{table*}

\subsection{Asymmetric Cryptography}
\label{sec:asycrypto}





\subsubsection{Modular Multiplication}
\label{sec:modular-multiplication}
Given three integers $x$, $y$ and $m$, this operation is to calculate 
$x*y \mod m$. 

\bheading{Naive and Karatsuba multiplications.}
Both OpenSSL and GnuPG implement two multiplication routines: naive multiplication and 
Karatsuba multiplication \cite{KaOf:62}. The routine is selected based on the size of 
the operands: the naive routine is adopted for multiplicands of small sizes, while Karatsuba 
routine is adopted for large multiplicands. 


Such implementation introduces control-flow side channels about the operands: Karatsuba 
routine is typically faster than the native routine. An adversary can measure the execution time
to infer the sizes of the operands, and then recover the secret key \cite{BrBo:05}. 



\subsubsection{Modular Exponentiation/Scalar Multiplication}
\label{sec:modular-exponentiation}
We consider the two operations together as they share similar implementations and 
vulnerabilities. Modular exponentiation is to calculate $x^y \mod m$, where $x$, $y$ and $m$ 
are three integers.
Scalar multiplication is to calculate $yx$ where $y$ is a scalar and $x$ is a point on the 
elliptic curve. The implementations of these two operations can reveal the secret key $y$ in
RSA and ElGamal, or secret scalar $y$ in ECC via side channels.



\bheading{Square-and-Multiply/Double-and-Add \cite{Go:98, HaMeVa:05}.}
The calculation of modular exponentiation is converted into a sequence of \texttt{SQUARE} 
and \texttt{MULTIPLY} operations. The binary representation of $y$ is denoted as
$y_{n-1}y_{n-2}...y_0$. Then starting from $n-1$ to 0, for each bit $y_i$, \texttt{SQUARE}
is called. If $y_i$ is 1, \texttt{MULTIPLY} is also called. Similarly, scalar multiplication 
is converted into a sequence of \texttt{PointDouble} and \texttt{PointAdd} based on each bit
$y_i$.


Such implementations are vulnerable to control-flow side-channel attacks: the execution of
\texttt{MULTIPLY} or \texttt{PointAdd} depends on bit $y_i$. By observing the traces of
\texttt{SQUARE} and \texttt{MULTIPLY} in modular exponentiation, or \texttt{PointDouble} and
\texttt{PointAdd} in scalar multiplication, an adversary can fully recover $y$. Successful 
attacks have been demonstrated against RSA in GnuPG via cache \textsc{Prime-Probe}
\cite{ZhJuRe:12, LiYaGe:15} and \textsc{Flush-Reload} \cite{YaFa:14} techniques, and against 
EdDSA via TLB side-channel \cite{GrRaBo:18}.

\bheading{Square-and-Multiply-always/Double-and-Add-always \cite{HaMeVa:05}.}
For modular exponentiation, this implementation always executes both \texttt{SQUARE} and
\texttt{MULTIPLY} operations for each bit. It selects the output of \texttt{SQUARE} if
$y_i$ is 0, and the output of \texttt{MULTIPLY} following \texttt{SQUARE} if $y_i$ is 1.
Similarly, Double-and-Add-always was proposed for scalar multiplication in ECC.



This implementation executes a fixed number of \texttt{SQUARE} and \texttt{MULTIPLY}
operations for modular exponentiation, or \texttt{PointDouble} and \texttt{PointAdd}
operations for scalar multiplication, defeating remote timing attacks.
However output selection still requires a secret-dependent branch, which is usually
smaller than one cache line. If it fits within the same cache line with preceding and succeeding
code, then it is not vulnerable to host-level attacks. However, Doychev and K{\"o}pf \cite{DoKo:17}
showed that for Libgcrypt, some compiler options can put this branch into separate cache lines, making
this implementation vulnerable to cache-based attacks. Gras et al. \cite{GrRaBo:18} showed that this
branch can put into separate pages, and the implementation is subject to TLB-based attacks.

\bheading{Sliding window \cite{BoCo:89}.}
For modular exponentiation, the exponent $y$ is represented as a sequence of windows $d_i$.
Each window starts and ends with bit $1$, and the window length cannot exceed a fixed parameter $w$.
So the value of any window is an odd number between $1$ and $2^w-1$. This method pre-computes
$x^v \mod m$ for each odd value $v \in [1, 2^w-1]$, and stores these results in a table indexed
by $i\in[0, (v-1)/2]$. Then it scans every window, squares and multiplies the corresponding entry
in the table.

Similarly, for scalar multiplication, the scalar $y$ is represented as a $w$-ary non-adjacent
form ($w$NAF), with each window value $d_i \in \{0,\pm1,...,\pm(2^{w-1}-1)\}$.
It first pre-computes the values of $\{1, 3, ..., 2^{w-1}-1\}x$, and stores them into a table. 
Then it scans each window, doubles and adds $d_jx$ if $d_j>0$ or subtracts $(-d_j)x$ otherwise.



Two types of vulnerabilities exist in such implementations. The first one is 
secret-dependent control flow: different routines will be called depending on
whether a window is zero. By monitoring the execution trace of those
branches, the adversary learns if each window is zero, and further recovers the secret.
Such attacks have been realized against RSA \cite{Pe:05, BeBrGe:17} and ECDSA
\cite{BeVaSm:14, VaSmYa:15, FaWaCh:16, AlBrFa:16, AlBrHa:19}.


The second one is secret-dependent data flow: the access location in the pre-computed table
is determined by each window value. By observing the access pattern, the adversary is able to
recover each window value. Attacks exploiting this vulnerability have been mounted against
RSA \cite{InGuIr:16}, ElGamal \cite{LiYaGe:15} and ECDSA \cite{BrHa:09}.




\bheading{Fixed window \cite{KaMeVa:96}.}
This method was designed to approach true constant-time implementation. Similar to sliding 
window, it also divides the secret $y$ into a set of windows, pre-computes the 
exponentiation or multiplication of each window value, and stores the results in a table. The 
differences are that the window size is fixed as $w$, and the table stores both odd and even
(including zero) values. It removes the critical control flow branch at the cost of more 
memory and slower run time.


To remove critical data flow, this approach can be combined with scatter-gather 
memory layout technique \cite{BrGrSe:06}, which stores the pre-computed values in
different cache lines instead of consecutive memory locations. Specifically,
each window value is stored across multiple cache lines, and each cache line stores
parts of multiple window values. When \texttt{MULTIPLY} or \texttt{PointAdd} is executed,
multiple cache lines are fetched to reconstruct the window value, hiding the
access pattern from the adversary. 


This implementation is still vulnerable to attacks \cite{YaGeHe:17} using cache bank,
the minimal data access unit in caches. Concurrent requests can be served
in parallel if they target different cache banks even in the same cache line, but have to be served
sequentially if they target the same cache bank. The timing difference between the
two cases enables the adversary to infer the window values accessed during the
gathering phase, and then recover the secret bits.



\bheading{Masked window.}
This approach was derived from fixed window implementation to further hide the cache bank
access patterns. The idea is to access all window values instead of just the one needed,
and then use a mask to filter out unused data. It performs a constant sequence of memory
accesses, and has been proven secure against different types of cache-based attacks \cite{DoKo:17}.

\bheading{Montgomery ladder \cite{Mo:87, JoYe:02}.}
This is a variation of Double-and-Add-always for scalar multiplication. It also represents $y$ in the
binary form and conducts both \texttt{PointAdd} and \texttt{PointDouble} functions for each bit,
irrespective of the bit value. The outputs of these functions are assigned to the 
intermediate variables determined by the bit value. A difference from Double-and-Add-always
is that in Montgomery ladder, the parameter of \texttt{PointDouble} is also determined by the 
bit value.

This implementation contains branches depending on the secret values. Yarom and Benger
\cite{YaBe:14} adopted cache \textsc{Flush-Reload} technique to identify the branch patterns and 
attack ECDSA in OpenSSL. 
Brumley and Tuveri \cite{BrTu:11} discovered that the implementation in OpenSSL 0.9.8 loops
from the most significant non-zero bit in $y$ to 0. So the number of iterations is proportional
to $log(y)$. This presents a vulnerability for remote timing attacks. 

\bheading{Branchless Montgomery ladder \cite{LaHaTu:16}.}
This approach replaces branches in Montgomery ladder with a function that uses bitwise logic to
swap two intermediate values only if the bit is 1, and thus removes the timing channel. However,
the implementations of \texttt{PointAdd} and \texttt{PointDouble} can still bring side channels.

First, OpenSSL adopts a lookup table to accelerate the square operation in the two functions.
The access pattern to the table can leak information about the secret in ECDH \cite{ShKiKw:18}.
Second, the modulo operation of $x \mod m$ in the two functions adopted the early exit implementation: if $x$ is 
smaller than $m$, $x$ is directly returned. This branch can be exploited by the adversary
to check whether $x$ is smaller than $m$, and then deduce secrets in EDH \cite{GeVaYa:17}. Third,
Kaufmann et al. \cite{KaPeVa:16} discovered that in Windows OS, the multiplication function 
of two 64-bit integers has an operand-dependent branch: if both operands 
have their 32 least significant bits equal to 0, then the multiplication is skipped and the 
result will be 0. This early exit branch was exploited to attack ECDH. 

\subsubsection{Modular Inverse}
\label{sec:modular-inversion}
This operation is to calculate the integer $x_m^{-1}$ such that $xx_m^{-1} \equiv 1 \mod m$.
It can also be used to check if two integers, $p$ and $q$, are co-prime.

\bheading{Binary Extended Euclidean Algorithm \cite{KaMeVa:96}.}
This approach uses arithmetic shift, comparison and subtraction to replace division. It is
particularly efficient for big integers, but suffers from control flow vulnerabilities
due to the introduction of operand-dependent branches.
Branch prediction \cite{AcGuSe:07} attacks was demonstrated to recover the value of $m$ in ECDSA and RSA. Page fault \cite{WeSpBo:18}
and cache \textsc{Flush-Reload} \cite{AlGaTa:18} techniques were adopted to attack the $gcd$ operation
in RSA key generation.


\bheading{Euclidean Extended Algorithm.}
This approach calculates quotients and remainders in each step without introducing 
secret-dependent branches. It is less efficient but secure against control flow side-channel attacks.



\subsection{Symmetric Cryptography}
\label{sec:aes}
In addition to asymmetric ciphers, symmetric ciphers, e.g., AES, can also be vulnerable 
to side-channel attacks.

\subsubsection{Substitution-Permutation}
This is a series of linked mathematical operations used in block ciphers.
It takes a block of the plaintext and the key as inputs, and applies several alternating 
``rounds'' of substitution boxes and permutation boxes to produce the 
ciphertext block. For AES, it consists of four basic operations: \texttt{ShiftRows}, 
\texttt{MixColumns}, \texttt{SubBytes} and \texttt{AddRoundKey}.

\bheading{T-box lookup.}
This approach converts the algebraic operations in each round into lookup table accesses. 
For AES, there are 8 pre-computed tables: $T_0$, $T_1$, $T_2$, $T_3$ are used in the first
9 rounds and $T_1^{10}$, $T_2^{10}$, $T_3^{10}$, $T_0^{10}$ are used in the final
round.
Each table contains 256 4-byte words. Then each round can be computed by updating the intermediate
states with certain words from the corresponding tables.



Since the accessed entries of the lookup tables are determined by the secret keys and 
plaintexts, an adversary can capture such access patterns to infer secrets.
Local cache attacks were proposed using \textsc{Prime-Probe} \cite{OsShTr:06},
\textsc{Evict-Time} \cite{OsShTr:06} and \textsc{Flush-Reload} \cite{GuBaKr:11, IrInEi:14} 
techniques. Remote timing attacks were proposed \cite{Be:05, BoMi:06} due to cache access collisions. 

\subsection{Post-Quantum Cryptography}
\label{sec:quantum}
Although post-quantum cryptography is secure against quantum computer based attacks, the
implementations of those algorithms can contain side-channel vulnerabilities, which can be 
attacked even by a conventional computer. 

\subsubsection{Distribution Sampling}
This operation is to sample an integer from a distribution. It is essential for BLISS
\cite{DuDuLe:13} to make the signature statistically independent of the secrets. However, an 
adversary can adopt side-channel attacks to recover the sampled data, and hence the secrets.


\bheading{Cumulative Distribution Table (CDT) Sampling \cite{Pe:10}.}
BLISS needs to sample blinding values from a discrete Gaussian distribution, and add them to
the signature. The CDT sampling approach pre-computes a table $\text{T}[i]=\mathbb{P}[x\leq i | x \sim D_{\sigma}]$. At the sampling phase, a random number $r$ is uniformly
chosen from $[0, 1)$, 
and the target $i$ is identified from T that satisfies $r \in [T[i-1], T[i])$. Some 
implementations adopt a guide table I to restrict the search space and accelerate the search 
process.

The access pattern to the two tables reveals information about the sampled values. An 
adversary can adopt cache \textsc{Flush-Reload} technique to recover the blinding values,
and further the secret key in BLISS \cite{BrHuLa:16, PeBrYa:17}.

\bheading{Rejection Sampling \cite{GePeVa:08}.}
This approach samples a bit from a Bernoulli distribution $B(\text{exp}(-x/2\sigma^2))$.
The implementation can bring side-channel opportunities for stealing the secret $x$:
(1) a lookup table $\text{ET}[i] = \text{exp}(-2^i/(2\sigma^2))$ is pre-computed to 
accelerate the bit sampling, causing a data flow vulnerability; (2) the sampling 
process needs to iterate over each secret bit and different branches will be executed for 
different bit values, producing a control flow vulnerability. 

Practical attacks were demonstrated exploiting those vulnerabilities. First, rejection 
sampling can replace CDT sampling for blinding value generation. An
adversary could utilize cache \cite{BrHuLa:16,PeBrYa:17} or branch \cite{EsFoGr:17} based attacks
to recover the sampled values in BLISS. Second, this approach can also be used to sample
random bits to probabilistically determine whether the blinding value is positive or negative,
and whether the signature should be accepted or rejected.
An adversary can infer the secret from this process via cache or branch
traces \cite{EsFoGr:17, TiWa:19}.

\subsubsection{Failure Rate Reduction}
Post-quantum schemes may have certain failure rate during encryption or decryption
due to its statistic nature. Thus it is necessary to devise methods to reduce the 
possibililty of failure.

\bheading{Error Correcting Code (ECC).}
This approach can significantly reduce the failure rate, but 
its implementation can reveal whether the ciphertext contains an error via timing channels: 
a ciphertext without an error is much faster to decode than one with errors. An adversary 
can exploit such information to recover the key \cite{DaTiVe:19}.

\subsubsection{Message Randomization}
Some post-quantum schemes require to randomize the message during encryption and decryption.
This process can also create side-channel vulnerabilities.

\bheading{Padding-Hash.}
In NTRU, encryption and decryption utilize hash functions to randomize the messages.\
However, the number of hash function calls highly depends on the input message. Thus, the 
total execution time of encryption or decryption will also differ for different inputs. By 
measuring such time information, an adversary is able to recover the secret input
\cite{SiWh:07}. 

\subsection{Cryptographic Protocol}
\label{sec:protocolpad}
Side-channel attacks were proposed to target the cryptographic protocols, specifically, the padding 
mechanism.

\subsubsection{RSA-PAD}
As introduced in \secref{sec:protocol-pad}, network protocols usually adopt RSA with PKCS. 
The padding mechanism can leak information about the plaintexts.

\bheading{Error message.}
In the handshake protocol in SSL 3.0, the receiver decrypts the message, and checks whether 
it is PKCS conforming. If so, she continues the handshake protocol. Otherwise she sends an
error message back to the sender and aborts the connection.
This message serves as a side channel to recover the plaintext. 
When the sender sends out a ciphertext, the adversary can intercept the message
and send a modified one to the receiver. From the receiver's response, the adversary learns if the 
first two bytes of the plaintext corresponding to the modified ciphertext are $0x00||0x02$
(valid PKCS conforming) or not
(invalid PKCS conforming). This can reduce the scope of the original plaintext. The adversary 
can repeat this process until the scope of plaintext is narrowed down to one single value. 

This vulnerability was discovered by Bleichenbacher \cite{Bl:98}, followed by variants of such attacks
\cite{KiPoFRo:03, BaFoKa:12}. Large-scale evaluations showed that it
still exists in many real-world websites, applications \cite{HaJuCr:18} and protocols \cite{FeGrSc:18}.




\bheading{Uniform response message.}
A common defense is to unify the responses for valid and invalid paddings:
if the decrypted message structure is not PKCS conforming, 
the receiver generates a random string as the plaintext, and 
performs all subsequent handshake computations on it. Thus, 
the adversary cannot distinguish valid ciphertexts from invalid ones based on the responses.


This implementation can still incur side-channel vulnerabilities. First, there can be other types of
messages that reveal the validation of padding format. For instance, in XML encryption, the message is 
encrypted using CBC mode, and the symmetric key is encrypted using RSA. 
Jager et al. \cite{JaScSo:12} discovered that the response of CBC encryption can leak whether the RSA padding 
is correct, enabling the adversary to recover the symmetric key. 



Second, there can be observable timing differences between valid and invalid padding cases.
Meyer et al. \cite{MeSoJu:14} discovered that in OpenSSL and JSSE, the receiver needs more
time to process the ill formatted message, due to generation of random numbers.
Jager et al. \cite{JaScSo:12} found that in XML encryption, invalid CBC decryption following
valid RSA decryption takes longer time than invalid RSA decryption.

Third, there can be control flow branches that depend on whether the message is PKCS conforming, 
such as error logging, data conversion and padding verification. They enable
a host-based adversary to monitor the execution trace as the oracle. Attacks based on control-flow
inference and cache \textsc{Flush-Reload} techniques \cite{ZhJuRe:14, XiLiCh:17, RoGiGe:19} were designed against different applications.


\subsubsection{CBC-MAC-PAD}
\label{sec:mac-cbc-pad}
The CBC-MAC padding scheme (\secref{sec:protocol-pad}) can also incur side-channel
vulnerabilities.

\bheading{Error message.}
When the receiver gets the ciphertext in CBC-MAC mode, she decrypts it and validates the padding format.
(1) If the format is invalid, she rejects the message, and sends a \emph{decryption\_failed} error message to the 
sender. Otherwise, she checks the MAC value. (2) If the MAC is incorrect, she returns a \emph{bad\_record\_mac} 
error message. (3) Otherwise, the message passes the validation and is accepted by the receiver.
These three conditions with three different responses create a side channel: an adversary can modify
the ciphertext and send it to the receiver for decryption. Based on the response, he can learn
whether the chosen ciphertext is decrypted into an incorrect padding. This oracle enables the
adversary to learn each byte of an arbitrary plaintext block. 

This vulnerability was first discovered by Vaudenay \cite{Va:02}. M{\"o}ller et al. \cite{MoDuKo:14}
designed an variant attack, POODLE, to compromise SSL 3.0. Evaluations on different applications
\cite{RiDu:10, DuRi:11} and protocols \cite{DePa:07, DePa:10} were conducted to show its severity
and pervasiveness.

\bheading{Uniform response message \cite{Mo:12}.}
This solution unifies the responses for both invalid padding format error and invalid MAC 
error, so the adversary cannot know whether the decrypted ciphertext has a valid padding or 
not.

Timing channels exist in this approach: if the format is invalid, the receiver only 
needs to perform simple operations on the very end of the ciphertext; otherwise,
she needs to further perform MAC validation throughout the whole ciphertext,
which takes more time. Attacks exploiting such timing discrepancy between these two
cases were demonstrated against TLS protocol in stunnel \cite{CaHiVa:03}, and DTLS in OpenSSL 
and GnuTLS \cite{AlPa:12}.


\bheading{Dummy MAC checking.}
This approach attempts to remove the timing channel by calling a dummy MAC validation even 
when the padding is incorrect. 
However, the calculation of MAC over a message $M$ depends on the message length: it takes more
time to compress longer $M$. Dummy MAC assumes the plaintext contains no padding, i.e.,
longest $M$. So an adversary can choose the ciphertext with short length of $M$ to distinguish whether
a real or dummy MAC is performed. AlFardan and Paterson \cite{FaPa:13} proposed lucky thirteen attacks, 
exploiting this vulnerability to recover plaintexts in TLS.




\bheading{Constant-time compression.}
Different strategies were designed to achieve constant-time compression implementations:
(1) dummy data can be appended to the padding for a maximum  (i.e., constant) number of MAC
compression operations;
(2) extra dummy compression operations can be added to make MAC validation constant time.

These implementations still contain secret-dependent control flows, making them 
vulnerable to host-level attacks. An adversary can obtain the padding validation results
via cache \textsc{Flush-Reload} \cite{IrLnEi:15}, \textsc{Prime-Probe} \cite{RoPaSh:18}
or control-flow inference \cite{XiLiCh:17} techniques.

\section{Summary of Countermeasures}
\label{sec:insights}

In this section, we summarize the cryptographic implementations from the perspective
of side-channel defenses.

From \secref{sec:crypto} we observe that one implementation
may be secure against one type of attack, but possibly vulnerable to another type of
attack. The side-channel resistance of one implementation highly depends
on the adversarial capabilities, i.e., type and granularity of observable side-channel
information. Table \ref{table:attacker's capability} summarizes the effectiveness of 
different countermeasure implementations under different types of attacks, which serves
as a reference for developers and users to select the optimal implementation based on
their security demands and threat models.

\begin{table*}[ht]
\centering
\caption{The effectiveness of implementations under different adversarial capabilities ($\blacksquare=$ vulnerable, $\square=$ secure)}
\label{table:attacker's capability}
\resizebox{\linewidth}{!}{
\begin{threeparttable}
\begin{tabular}{|c|c|c|c|c|c|c|c|c|c|c|}
  \hline
  \multirow{2}{*}{\textbf{Category}} & \multirow{2}{*}{\textbf{Operation}} & \multirow{2}{*}{\textbf{Implementation}} & \multicolumn{2}{|c|}{\textbf{Network-level}} & \multicolumn{5}{|c|}{\textbf{Host-level}} & \multirow{2}{*}{\textbf{Index}} \\
  \cline{4-10}
  & & & \textbf{Timing} & \textbf{Message} & \textbf{Page} & \textbf{Cacheline} & \textbf{CacheBank} & \textbf{Branch} & \textbf{$\mu$-ops}& \\
  \hline
  \hline

  & Modular Multiplication & Basic and Karatsuba multiplication & $\blacksquare$ & $\square$ & $\blacksquare$ & $\blacksquare$ & $\blacksquare$ & $\blacksquare$ & $\blacksquare$ & 1.1.1\\
  \cline{2-11}

  & \multirow{6}{*}{Modular exponentiation} & Square(Double)-and-Multiply(Add) & $\blacksquare$ & $\square$ & $\blacksquare$ & $\blacksquare$ & $\blacksquare$ & $\blacksquare$ & $\blacksquare$ & 1.2.1\\
  \cline{3-11}

  & \multirow{6}{*}{Scalar multiplication} & Square(Double)-and-Multiply(Add)-always & $\square$ & $\square$ & $\blacksquare$ & $\blacksquare$ & $\blacksquare$ & $\blacksquare$ & $\blacksquare$ & 1.2.2\\
  \cline{3-11}
  & & Sliding window & $\blacksquare$ & $\square$ & $\blacksquare$ & $\blacksquare$ & $\blacksquare$ & $\blacksquare$ & $\blacksquare$ & 1.2.3\\
  \cline{3-11}
  & & Fixed window & $\square$ & $\square$ & $\square$ & $\square$ & $\blacksquare$ & $\square$ & $\square$ & 1.2.4 \\
  \cline{3-11}
  Asymmetric & & Masked window& $\square$ & $\square$ & $\square$ & $\square$ & $\square$ & $\square$ & $\square$ & 1.2.5 \\
  \cline{3-11}
  Cryptography & & Montgomery ladder & $\square$ & $\square$ & $\blacksquare$ & $\blacksquare$ & $\blacksquare$ & $\blacksquare$ & $\blacksquare$ & 1.2.6 \\
  \cline{3-11}
  & & Branchless montgomery ladder & $\square$ & $\square$ & $\square$ & $\square$ & $\square$ & $\square$ & $\square$ & 1.2.7 \\
  \cline{2-11}

  & \multirow{2}{*}{Modular inverse} & Binary Extended Euclidean Algorithm & $\blacksquare$ & $\square$ & $\blacksquare$ & $\blacksquare$ & $\blacksquare$ & $\blacksquare$ & $\blacksquare$ & 1.3.1 \\
  \cline{3-11}
  & & Extended Euclidean Algorithm & $\square$ & $\square$ & $\square$ & $\square$ & $\square$ & $\square$ & $\square$ & 1.3.2 \\
  \cline{2-11}

  & RSA, ElGamal, & Key blinding & $\square$ & $\square$ & $\blacksquare$ & $\blacksquare$ & $\blacksquare$ & $\blacksquare$ & $\blacksquare$ & -- \\
  \cline{3-11}
  & ECDH, ECDSA & Plaintext/ciphertext blinding & $\square$ & $\square$ & $\square$ & $\square$ & $\square$ & $\square$ & $\square$ & -- \\
  \hline

  Symmetric & \multirow{2}{*}{Substitution-Permutation} & T-box lookup & $\blacksquare$ & $\square$ & $\square$ & $\blacksquare$ & $\blacksquare$ & $\square$ & $\square$ &  2.1.1 \\
  \cline{3-11}
  Cryptography & & AES-NI & $\square$ & $\square$ & $\square$ & $\square$ & $\square$ & $\square$ & $\square$ & -- \\
  \hline

  & \multirow{2}{*}{Distribution Sampling} & Cumulative Distribution Table Sampling & $\square$ & $\square$ & $\blacksquare$ & $\blacksquare$ & $\blacksquare$ & $\blacksquare$ & $\blacksquare$ & 3.1.1 \\
  \cline{3-11}
  Post-Quantum & & Rejection Sampling & $\square$ & $\square$ & $\blacksquare$ & $\blacksquare$ & $\blacksquare$ & $\blacksquare$ & $\blacksquare$ & 3.1.2 \\
  \cline{2-11}
  Cryptography  & Failure Rate Reduction & Error Correcting Code & $\blacksquare$ & $\square$ & $\blacksquare$ & $\blacksquare$ & $\blacksquare$ & $\blacksquare$ & $\blacksquare$ & 3.2.1 \\
  \cline{2-11}
  & Message Randomization & Padding-Hash & $\blacksquare$ & $\square$ & $\blacksquare$ & $\blacksquare$ & $\blacksquare$ & $\blacksquare$ & $\blacksquare$ & 3.3.1\\
  \cline{2-11}
  \hline

  & \multirow{2}{*}{RSA-PAD} & Error message & $\blacksquare$ & $\blacksquare$ & $\blacksquare$ & $\blacksquare$ & $\blacksquare$ & $\blacksquare$ & $\blacksquare$ & 4.1.1 \\
  \cline{3-11}
  & & Uniform response message & $\blacksquare$ & $\square$ & $\blacksquare$ & $\blacksquare$ & $\blacksquare$ & $\blacksquare$ & $\blacksquare$ & 4.1.2 \\
  \cline{2-11}
  Protocol & \multirow{4}{*}{CBC-MAC-PAD} & Error message & $\blacksquare$ & $\blacksquare$ & $\blacksquare$ & $\blacksquare$ & $\blacksquare$ & $\blacksquare$ & $\blacksquare$ & 4.2.1\\
  \cline{3-11}
  Padding & & Uniform response message & $\blacksquare$ & $\square$ & $\blacksquare$ & $\blacksquare$ & $\blacksquare$ & $\blacksquare$ & $\blacksquare$ & 4.2.2 \\
  \cline{3-11}
  & & Dummy MAC checking & $\blacksquare$ & $\square$ & $\blacksquare$ & $\blacksquare$ & $\blacksquare$ & $\blacksquare$ & $\blacksquare$ & 4.2.3 \\
  \cline{3-11}
  & & Constant-time compression & $\square$ & $\square$ & $\blacksquare$ & $\blacksquare$ & $\blacksquare$ & $\blacksquare$ & $\blacksquare$ & 4.2.4 \\
  \hline

  \end{tabular}
  \end{threeparttable}}
\vspace{-5pt}
\end{table*}

The goal of side-channel defenses is to eliminate the correlation between the application's
secrets and runtime behaviors. This can be achieved by either \emph{unifying} or \emph{randomizing}
the side-channel information. Different approaches may share common routines and features. Below
we abstract the key features of these techniques.

\subsection{Side-channel Information Unification}

\subsubsection{Remove Control Flow}
Control flow vulnerabilities exist when different values of the secret $s$ lead to different
code paths distinguishable by the adversary (Figure \ref{fig:insecure}).
Two strategies can be used to remove such control flow.

\begin{figure}[t]
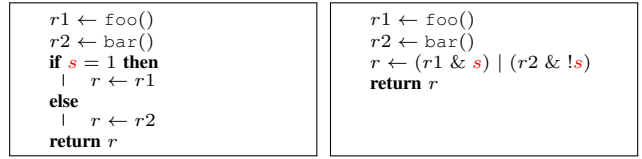

  \centering
  \removelatexerror
  \begin{minipage}[t][][b]{.49\linewidth}
    \begin{algorithm*}[H]
      \scriptsize
      \SetAlgoLined
      \Indp
      \Indm
      \SetKwProg{myalg}{function}{}{end}
        $r1\gets \texttt{foo}()$ \\
        $r2\gets \texttt{bar}()$ \\
        \eIf{$\textcolor{red}{s}=1$} {
          $r\gets r1$
        } {
          $r\gets r2$
        }
      \KwRet{$r$} 
    {}
    \end{algorithm*}
  \captionsetup{labelformat=empty}
  \subcaption{\scriptsize{AlwaysExecute-ConditionalSelect}}
  \label{fig:execute-always}
  \end{minipage} \hfill
  \begin{minipage}[t][][b]{.49\linewidth}
    \begin{algorithm*}[H]
      \scriptsize
      \SetAlgoLined
      \Indp
      \Indm
      \SetKwProg{myalg}{function}{}{end}
        $r1\gets \texttt{foo}()$ \\
        $r2\gets \texttt{bar}()$ \\
        $r\gets (r1 \;\& \;\textcolor{red}{s}) \;| \;(r2 \;\& \;!\textcolor{red}{s})$ \\
      \KwRet{$r$} \\
        \fontsize{20}{22}\selectfont $\;$ \\
    {}
    \end{algorithm*}
  \captionsetup{labelformat=empty}
  \subcaption{\scriptsize{AlwaysExecute-BitwiseSelect}}
  \label{fig:bitwise-logic}
  \end{minipage}
  \caption{Remove control flow vulnerability}
  \label{fig:remove-control-flow}
  \vspace{-10pt}
\end{figure}

\bheading{AlwaysExecute-ConditionalSelect.}
Each possible routine is always executed regardless of the condition. Based on the secret
value, the correct result is assigned to the return variable (Figure \ref{fig:execute-always}).
This technique is adopted in modular exponentiation (Square-and-Multiply-always), scalar
multiplication (Double-and-Add-always) and CBC-MAC-PAD (dummy MAC checking).

This solution is effective against network-level attacks. However, control flow
in result selection can still be observed by an adversary via host-level attacks. Besides, if the
values of each code path are pre-computed and stored in memory, the adversary can also infer the
secret via data flow, exemplified by sliding window implementations in modular exponentiation and
scalar multiplication.



\bheading{AlwaysExecute-BitwiseSelect.}
This strategy performs all possible computations, and then selects the correct result using bitwise
operations of secret $s$ (Figure \ref{fig:bitwise-logic}). This introduces no branches
or access patterns. The branchless Montgomery ladder algorithm adopts this solution for
constant-time conditional swap in scalar multiplication.

\RestyleAlgo{boxed}
\SetAlgoLined
\LinesNumberedHidden

\begin{figure}[t]
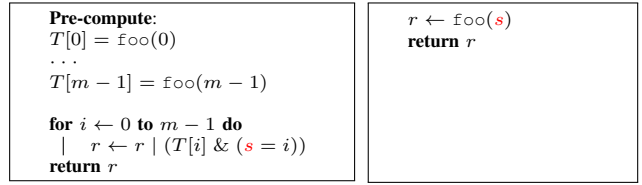

  \centering
  \removelatexerror
  \begin{minipage}[t][][b]{.55\linewidth}
    \begin{algorithm*}[H]
      \scriptsize
      \SetAlgoLined
      \Indp
      \Indm
      \SetKwProg{myalg}{function}{}{end}
        \textbf{Pre-compute}: \\
        $T[0] = \texttt{foo}(0)$ \\
        $\cdot\cdot\cdot$ \\
        $T[m-1] = \texttt{foo}(m-1)$ \\
        $\;$ \\
        \For{$i \gets 0$ \textbf{\emph{to}} $m-1$ } {
          $r\gets r \;|\; (T[i] \;\& \;(\textcolor{red}{s}=i))$
        }
      \KwRet{$r$}
    {}
    \end{algorithm*}
  \captionsetup{labelformat=empty}
  \subcaption{\scriptsize{AlwaysAccess-BitwiseSelect}}
  \label{fig:masked-access}
  \end{minipage} \hfill
  \begin{minipage}[t][][b]{.43\linewidth}
    \begin{algorithm*}[H]
      \scriptsize
      \SetAlgoLined
      \Indp
      \Indm
      \SetKwProg{myalg}{function}{}{end}
      $r\gets \texttt{foo}(\textcolor{red}{s})$ \\
      \KwRet{$r$} \\
        \fontsize{20}{48}\selectfont $\;$ \\
    {}
    \end{algorithm*}
  \captionsetup{labelformat=empty}
  \subcaption{\scriptsize{On-the-fly calculation}}
  \label{fig:on-the-fly}
  \end{minipage}
  \caption{Remove data flow vulnerability}
  \label{fig:remove-data-flow}
  \vspace{-10pt}
\end{figure}

\subsubsection{Remove Data Flow}
Data flow vulnerabilities exist when different values of the secret $s$ cause different
memory accesses that can be observed by the adversary (Figure \ref{fig:memory-traffic}).
Two strategies can remove such data flow.

\bheading{AlwaysAccess-BitwiseSelect.}
This method accesses all critical locations, and selects the correct value based on the bitwise
operation (Figure \ref{fig:masked-access}). It is adopted in masked window modular
exponentiation and scalar multiplication.

\bheading{On-the-fly calculation.}
We can calculate the value every time it is used instead of pre-computing all values and
storing them into a table, particularly when the calculation is not
complex and does not introduce secret-dependent control flows. Branchless Montgomery ladder
adopts this method in the square operation of scalar multiplication.


\subsection{Side-channel Information Randomization}

\subsubsection{Cryptographic Blinding}
This solution does not focus on specific primitives or operations. Instead, it 
improves the high-level asymmetric cipher algorithms. There are generally two
types of blinding techniques.


\bheading{Key blinding.}
A random factor is blended into the secret key, but the original key and the randomized key
generate the same cryptographic result. The adversary can only obtain the randomized key
via side-channel attacks, which is useless without knowing the blended random factor.


For ECDSA and ECDH, the randomized key is $k+sr$ where $r$ is a random number and $s$ is
the group order. The scalar multiplication generates $(k+sr)G$, the same as $kG$ \cite{Co:99}.
For RSA and ElGamal, the randomized key is $d+r\phi(n)$ where $r$ is a random number and
$\phi$ is the Euler's totient function. The decryption gives $c^{d+r\phi(n)} \mod n$, the same
as $c^d \mod n$. In both cases, the true value of $k$ is hidden from side-channel adversaries.

\bheading{Plaintext/ciphertext blinding.}
This approach randomizes the plaintexts or ciphertexts, adaptively chosen by the adversary.
The randomized texts cause the adversary to recover a wrong key via
side-channel analysis. This solution works only if correct ciphertexts can be produced from
randomized plaintexts and vice versa. 


For ECDSA and ECDH, we can choose a random point $R$ and use $G'=G+R$ in computation. The adversary
cannot recover $k$ from the side-channel observation without the knowledge of $R$ \cite{Co:99}, but
we can easily reproduce the correct result $kG$ by subtracting $kR$ from $kG'$. For RSA and
ElGamal, we can generate a random value $r$, and replace $c$ with $c*r^e$. Now the decryption
process is randomized to be $(c*r^e)^d \mod n = c^d * r^{ed} \mod n$.
To get $c^d \mod n$ we can simply multiply the result by $r^{-1}$, as $r^{ed} * r^{-1} \equiv 1 \mod n$.




\section{Evaluation of Cryptographic Libraries}
\label{sec:history}

From a practical perspective, we review, analyze and evaluate the development of side-channel
attacks and defenses in two commonly used cryptographic libraries: OpenSSL and GNU Crypto (GnuPG,
Libgcrypt and GnuTLS). We collect side-channel related history (1999 -- 2019) from Common
Vulnerabilities and Exposures (CVE), changelogs and commit messages, and source code. Table
\ref{table:crypto-history} shows the evolution of the libraries. The full history
is tabulated in Appendix \ref{sec:release-lib}. 


\begin{table*}[h]
\caption{Evolution of cryptographic libraries for side-channel activities}
\label{table:crypto-history}
\begin{subtable}[h]{\linewidth}
\centering
\resizebox{0.95\linewidth}{!}{
\begin{tabular}{|c|c|c|c|c|c|c|}
  \hline
  \textbf{Date} & \textbf{Version} & \textbf{Vulnerable Operations} & \textbf{Vulnerable Implementation} & \textbf{CVE} & \textbf{Countermeasures} & \textbf{Index} \\
  \hline
  \hline

  2001/07/09 & 0.9.6b & RSA-PAD & Uniform error message & & Fix bugs & \\
  \hline
  2003/02/19 & 0.9.6i, 0.9.7a & CBC-MAC-PAD & Uniform error message & CVE-2003-0078 & Dummy checking for TLS & 4.2.2.a \\
  \hline
  \multirow{2}{*}{2003/04/10} & \multirow{2}{*}{0.9.6j, 0.9.7b} & Modular multiplication & Basic and Karatsuba multiplication & CVE-2003-0147 & RSA blinding & 1.1.1.a \\
  \cline{3-7}
  & & RSA-PAD & Uniform error message & CVE-2003-0131 & Uniform version error message & 4.1.1.b \\
  \hline
  2005/07/05 & 0.9.8 & \multirow{2}{*}{Modular exponentiation} & \multirow{2}{*}{Sliding window} & & \multirow{2}{*}{Fixed window} & \\
  \cline{1-2}
  2005/10/11 & 0.9.7h &  &  & &  & \\
  \hline
  2007/10/11 & 0.9.8f & Modular inversion & Binary Extended Euclidean Algorithm &  & Euclidean Extended Algorithm & \\
  \hline
  2011/09/06 & 1.0.0e & \multirow{2}{*}{Scalar multiplication} & \multirow{2}{*}{Montgomery ladder} & \multirow{2}{*}{CVE-2011-1945} & \multirow{2}{*}{Make the bit length of scalar constant} & 1.2.6.b \\
  \cline{1-2}
  \multirow{3}{*}{2012/01/04} & 0.9.8s & & & & & \\
  \cline{2-7}
  & \multirow{2}{*}{0.9.8s, 1.0.0f} & \multirow{2}{*}{CBC-MAC-PAD} & Uniform error message & CVE-2011-4108 & Dummy checking for DTLS & 4.2.2.b \\
  \cline{4-7}
  & & & Padding data initialization & CVE-2011-4576 & Fix bugs & \\
  \hline
  2012/03/12 & 0.9.8u, 1.0.0h & RSA-PAD  (PKCS \#7 and CMS) & Error message & CVE-2012-0884 & Uniform error message and dummy checking & \\
  \hline
  \multirow{2}{*}{2012/03/14} & \multirow{2}{*}{1.0.1} & Scalar multiplication & Sliding window & & Masked window & \\
  \cline{3-7}
  & & Substitution-Permutation & T-box lookup & & AES-NI support & \\
  \hline
  2013/02/05 & 0.9.8y, 1.0.0k, 1.0.1d & CBC-MAC-PAD & Dummy MAC checking & CVE-2013-0169 & Dummy data padding & 4.2.3.a \\
  \hline
  2014/04/07 & 1.0.1g & \multirow{2}{*}{Scalar multiplication} & \multirow{2}{*}{Montgomery ladder} & \multirow{2}{*}{CVE-2014-0076} & \multirow{2}{*}{Branchless Montgomery ladder} & 1.2.6.a \\
  \cline{1-2}
  2014/06/05 & 0.9.8za, 1.0.0m & & & & & \\
  \hline
  2014/10/15 & 0.9.8zc, 1.0.0o, 1.0.1j & CBC-MAC-PAD & Error message & CVE-2014-3566 & Disable fallback of SSLv3.0 & 4.2.1.b \\
  \hline
  \multirow{4}{*}{2016/01/28} & \multirow{4}{*}{1.0.1r, 1.0.2f} & \multirow{4}{*}{RSA-PAD} & \multirow{4}{*}{Uniform error message} & CVE-2015-3197 & \multirow{4}{*}{Disable SSLv2 ciphers} & \multirow{4}{*}{4.1.2.b}\\
  & & & & CVE-2016-0703 &  & \\
  & & & & CVE-2016-0704 &  & \\
  & & & & CVE-2016-0800 &  & \\
  \hline
  \multirow{5}{*}{2016/03/01} & \multirow{5}{*}{1.0.1s, 1.0.2g} & \multirow{4}{*}{RSA-PAD} & \multirow{4}{*}{Uniform error message} & CVE-2015-3197 & \multirow{4}{*}{Disable SSLv2 protocols} & \multirow{4}{*}{4.1.2.b} \\
  & & & & CVE-2016-0703 &  & \\
  & & & & CVE-2016-0704 &  & \\
  & & & & CVE-2016-0800 &  & \\
  \cline{3-7}
  & & Modular exponentiation & Fixed window& CVE-2016-0702 & Masked window & 1.2.4.a \\
  \hline
  2016/05/03 & 1.0.1t, 1.0.2h & CBC-MAC-PAD (AES-NI) & Dummy data padding & CVE-2016-2107 &Fix bugs & \\
  \hline
  2016/09/22 & 1.0.1u, 1.0.2i & Modular exponentiation & Fixed window & CVE-2016-2178 & Fix bugs & \\
  \hline
  \multirow{4}{*}{2018/08/14} & \multirow{3}{*}{1.0.2p, 1.1.0i} & Scalar multiplication & Branchless Montgomery ladder & & On-the-fly calculation to replace lookup table & \\
  \cline{3-7}
  & & Modular inversion & Binary Greatest Common Divisor & CVE-2018-0737 & Extended Euclidean Algorithm & 1.3.1.c \\
  \cline{3-7}
  & & Modulo & Early exit &  & ECDSA and DSA blinding & \\
  \cline{2-7}
  & 1.1.0i & Scalar multiplication & Sliding window & & Branchless Montgomery ladder & \\
  \hline  
  \multirow{6}{*}{2018/09/11} & \multirow{6}{*}{1.1.1} & \multirow{4}{*}{Scalar multiplication} & \multirow{2}{*}{Branchless Montgomery ladder} & & Differential addition-and-doubling & \\
  \cline{6-6}
  & & & & & Coordinate blinding &  \\
  \cline{4-7}  
  & & & Masked window & & Branchless Montgomery ladder & \\
  \cline{4-7}  
  & & & Sliding window & & Branchless Montgomery ladder & \\
  \cline{3-7}
  & & \multirow{2}{*}{Modular inversion} & \multirow{2}{*}{Extended Euclidean Algorithm} & & Implementing new constant-time function for EC & \\
  \cline{6-6}
  & & & & & Input blinding &  \\
  \hline
  \multirow{3}{*}{2018/11/20} & 1.0.2q& Scalar multiplication (P-384) & Sliding window & CVE-2018-5407 &  Branchless Montgomery ladder & 1.2.3.d \\
  \cline{2-7}
  & \multirow{2}{*}{1.0.2q, 1.1.0j, 1.1.1a} & DSA sign setup & Space preallocation & CVE-2018-0734 & Fix bugs & \\
  \cline{3-7}
  & & Scalar multiplication & Space preallocation & CVE-2018-0735 & Fix bugs & \\
  \hline
  \multirow{2}{*}{2019/02/26} & 1.0.2r & CBC-MAC-PAD & Protocol error handling & CVE-2019-1559 & Fix bugs &  \\
  \cline{2-7}
  & 1.1.1b & Modular inversion (EC) & Binary Extended Euclidean Algorithm &  & EC-specific inversion function with input blinding &  \\
  \hline
\end{tabular}}
\caption{OpenSSL}
\label{table:openssl-history}
\end{subtable}

\begin{subtable}[h]{\linewidth}
\centering
\resizebox{0.95\linewidth}{!}{
\begin{tabular}{|c|c|c|c|c|c|c|}
  \hline
  \textbf{Date} & \textbf{Version} & \textbf{Vulnerable Operations} & \textbf{Vulnerable Implementation} & \textbf{CVE} & \textbf{Countermeasures} & \textbf{Index} \\
  \hline
  \hline

  2006/09/08 & T1.4.3 & \multirow{2}{*}{RSA-PAD} & \multirow{2}{*}{Error Message} & & \multirow{2}{*}{Uniform error message} & \\
  \cline{1-2}
  2006/09/21 & T1.5.1 & & & & & \\
  \hline
  2011/06/29 & L1.5.0 & Substitution-permutation & T-box lookup & & AES-NI support & \\
  \hline
  2012/01/06 & T3.0.11 & CBC-MAC-PAD & Uniform error message & CVE-2012-0390 & Dummy checking for DTLS & 4.2.2.b \\
  \hline
  2013/02/04 & T2.12.23, T3.0.28, T3.1.7 & CBC-MAC-PAD & Dummy MAC checking & CVE-2013-1619 & Dummy data padding & 4.2.3.a \\
  \hline
  2013/07/25 & P1.4.14,  L1.5.3 & Modular exponentiation & Square-and-Multiply & CVE-2013-4242 & Square-and-Multiply-always & 1.2.1.a \\
  \hline
  2013/12/16 & L1.6.0 & Modular exponentiation & Square-and-Multiply & CVE-2013-4242 & Square-and-Multiply-always & 1.2.1.a \\
  \hline
  2013/12/18 & P1.4.16 & Modular multiplication & Basic and Karatsuba multiplication & CVE-2013-4576 & Exponentiation blinding & \\
  \hline
  2014/08/07 & L1.5.4 & Modular multiplication & Basic and Karatsuba multiplication & CVE-2014-5270 & Exponentiation blinding & \\
  \hline
  \multirow{2}{*}{2015/02/27} & \multirow{2}{*}{P1.4.19, L1.6.3} & Modular multiplication & Basic and Karatsuba multiplication & CVE-2014-3591 & ElGamal Blinding & \\
  \cline{3-7}
  & & Modular exponentiation & Sliding window & CVE-2015-0837 & Remove control flow of multiply operation & 1.2.1.b \\
  \hline
  2016/02/09 & L1.6.5 & \multirow{2}{*}{Scalar multiplication} & \multirow{2}{*}{Sliding window} & \multirow{2}{*}{CVE-2015-7511} & \multirow{2}{*}{Double-and-Add-always} & \\
  \cline{1-2}
  \multirow{2}{*}{2016/02/18} & \multirow{2}{*}{L1.5.5} & & & & & \\
  \cline{3-7}
  & & \multirow{2}{*}{Modular multiplication} & \multirow{2}{*}{Basic and Karatsuba multiplication} & \multirow{2}{*}{CVE-2014-3591} & \multirow{2}{*}{ElGamal Blinding} & \\
  \cline{1-2}
  \multirow{3}{*}{2016/04/15} & \multirow{3}{*}{L1.7.0} & & & & & \\
  \cline{3-7}
  & & Modular exponentiation & Sliding window & CVE-2015-0837 & Remove control flow of multiply operation & 1.2.1.b \\
  \cline{3-7}
  & & Scalar multiplication & Sliding window & CVE-2015-7511 & Double-and-Add-always & \\
  \hline
  2017/06/29 & L1.7.8 & \multirow{3}{*}{Modular exponentiation} & \multirow{3}{*}{Sliding window} & \multirow{3}{*}{CVE-2017-7526} & \multirow{3}{*}{RSA exponentiation blinding} & \multirow{3}{*}{1.2.3.b}\\
  \cline{1-2}
  2017/07/18 & L1.8.0 & & & & & \\
  \cline{1-2}
  2017/07/19 & P1.4.22 & & & & & \\
  \hline
  2017/08/27 & L1.7.9, L1.8.1 & Scalar multiplication & Branchless montgomery ladder & CVE-2017-0379 & Input validation & 1.2.7.b \\
  \hline
  2018/06/13 & L1.7.10, L1.8.3 & Modulo & Early exit & CVE-2018-0495 & ECDSA blinding & \\
  \hline
  \multirow{3}{*}{2018/07/16} & \multirow{3}{*}{T3.3.30, T3.5.19, T3.6.3} & \multirow{3}{*}{CBC-MAC-PAD} & \multirow{3}{*}{Pseudo constant time} & CVE-2018-10844 & \multirow{2}{*}{New variant of pseudo constant time} & \\
  \cline{5-5}
  & & & & CVE-2018-10845 & \multirow{2}{*}{(Not fully mitigate the vulnerability))} & 4.2.4.c \\
  \cline{5-5}
  & & & & CVE-2018-10846 & & \\
  \hline
  2018/12/01 & T3.6.5 & RSA-PAD & Pseudo constant time & CVE-2018-16868 & Hide access pattern and timing & 4.1.2.d\\
  \hline

  \end{tabular}}
  \caption{GNU crypto (For the version column, P: GnuPG; L: Libgcrypt; T: GnuTLS)}
  \label{table:gnupg-history}
  \end{subtable}
\end{table*}

\subsection{Vulnerability severity}
\label{sec:vul-severe}


We examine the severity and practicality of side-channel attacks as well as the attention
developers paid to them. We establish the measurements for these threats and compare them
with other vulnerability categories.

We adopt the Common Vulnerability Scoring System (CVSS)\footnote{The latest CVSS version 
is v3.0. In this paper, we adopt CVSS v2.0, as old vulnerabilities were not 
assigned CVSS v3.0 scores.}, an industry standard, to assess each CVE. The 
score ranges from 0 (least severe) to 10 (most severe). 
We consider the \emph{Base} score that well represents the inherent quality of a 
vulnerability. It comprises two sub-scores, \emph{Exploitability} that defines 
the difficulty to attack the software and \emph{Impact} that defines the 
level of damage to certain properties of the software under a successful
attack. Appendix \ref{sec:cvss-cal} details computation of these scores.

\begin{figure}[t]
\centerline{\mbox{\includegraphics[width=\linewidth]{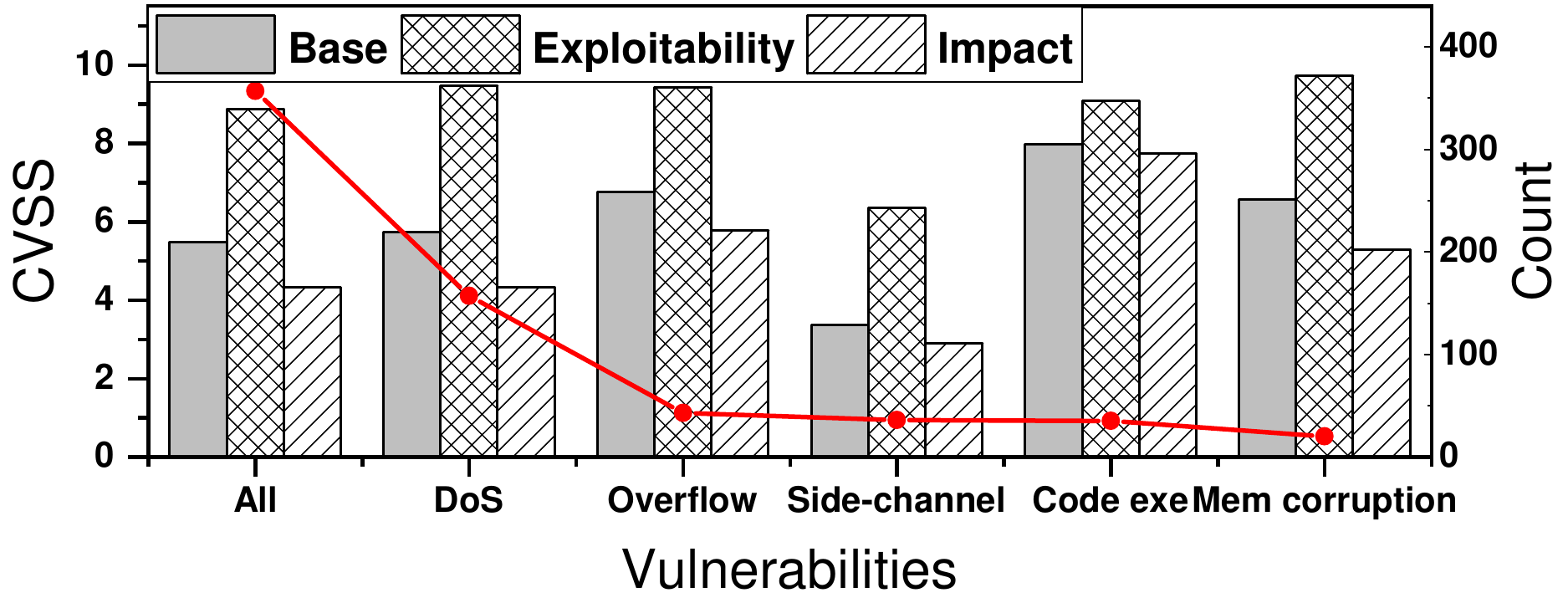}}}
\vspace{-5pt}
\caption{CVSS score of different software vulnerabilities.}
\label{fig:cve-1}
\vspace{-10pt}
\end{figure}

For OpenSSL and GNU Crypto, the top vulnerabilities are denial-of-service, arbitrary 
code execution, buffer overflow, and memory corruption.
Figure \ref{fig:cve-1} compares the average scores and quantities of these vulnerability
categories\footnote{There are some mistakes in CVEs: (1) all 
side-channel vulnerabilities should only have partial confidentiality impact, while 
CVE-2003-0131, CVE-2013-1619 and CVE-2018-16868 were also assigned partial integrity or 
availability impact. (2) CVE-2018-10844, CVE-2018-10845 and CVE-2018-10846 should have local 
access vector, but they were assigned network access vector. We corrected them 
in our analysis.}. We observe that \textbf{\emph{side-channel vulnerabilities are regarded less severe 
than other types}} due to lower \emph{Exploitability} and \emph{Impact} sub-scores.
Side-channel attacks usually require stronger adversarial capabilities, in-depth knowledge about
underlying platforms, and a large amount of attack sessions, but only cause partial
confidentiality breach as they leak (part of) keys or plaintexts. In contrast, other vulnerabilities
may be exploited by less-experienced attackers, but enable them to execute arbitrary code or disable the services entirely.

\begin{figure}[t]
  \centering
  \begin{subfigure}{0.55\linewidth}
    \centering
    \includegraphics[width=\linewidth]{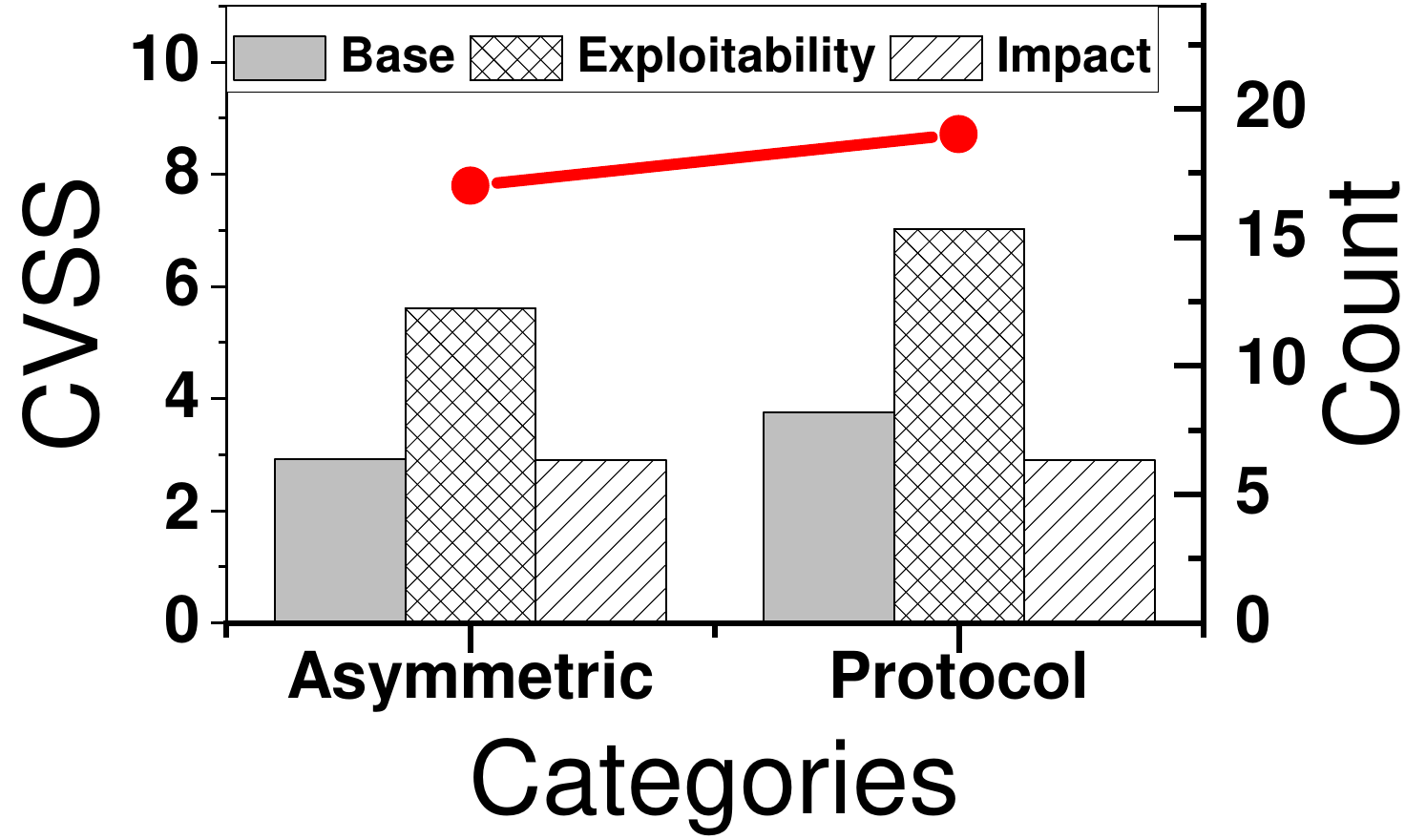}
    \caption{\scriptsize{CVSS Scores}}
    \label{fig:cve-2}
  \end{subfigure}%
  \begin{subfigure}{0.45\linewidth}
    \centering
    \includegraphics[width=\linewidth]{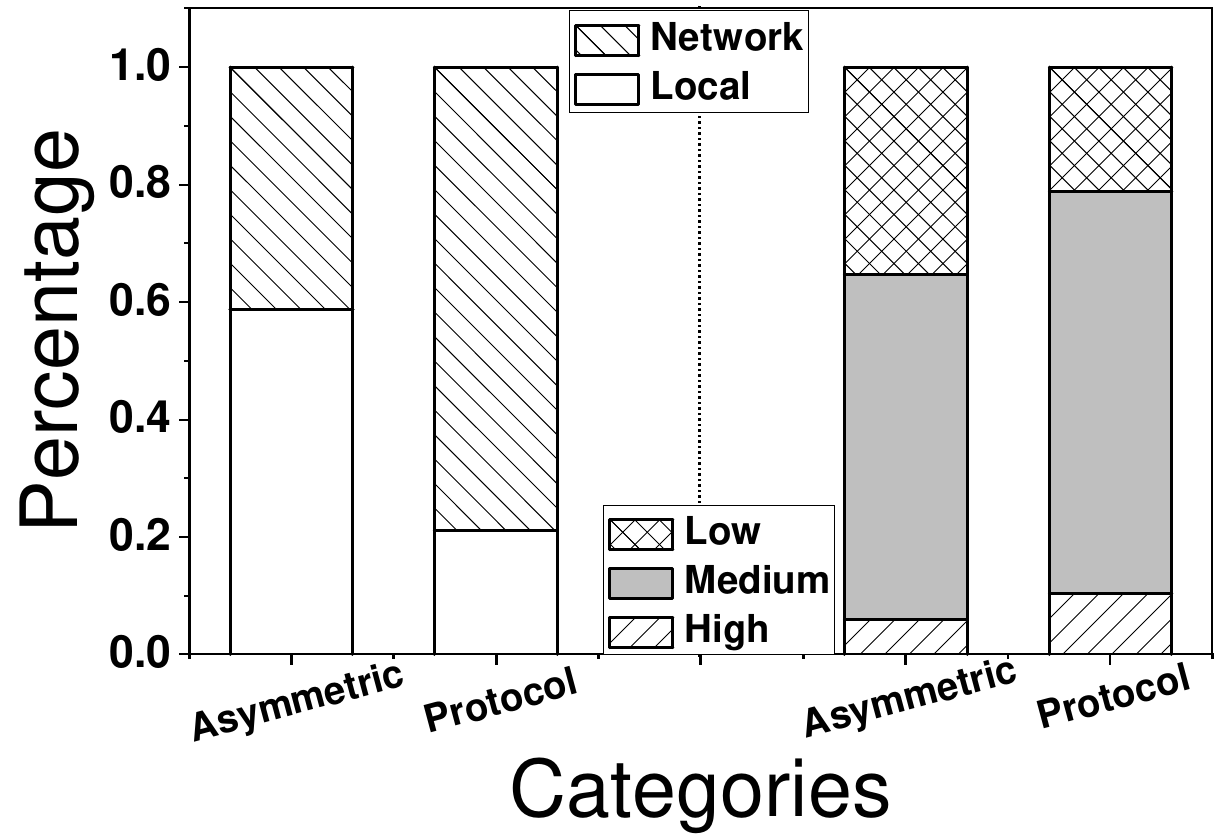}
    \caption{\scriptsize{Access vector and complexity}}
    \label{fig:cve-4}
  \end{subfigure}
  \caption{Side-channel vulnerabilities in different operations}
  \label{fig:access-vector-complexity}
\vspace{-10pt}
\end{figure}


Next we compare side-channel vulnerabilities in two categories: asymmetric ciphers
and protocol padding (we skip symmetric ciphers and post-quantum cryptography as
fewer vulnerabilities were identified in their libraries). Figure \ref{fig:cve-2} shows the 
average CVSS scores of each category. We observe that \textbf{\emph{vulnerabilities in protocol
padding are generally more severe that those in asymmetric ciphers}} due to higher
\emph{Exploitability}.
The underlying reason is that \emph{Exploitability} is determined by access vector and access
complexity (Appendix \ref{sec:cvss-cal}). Figure \ref{fig:cve-4} 
shows the breakdown of these two factors in each category. For access vector, 
network vector and local vector are neck and neck for vulnerabilities in asymmetric 
ciphers, but the former dominates access vectors of padding oracle attacks, rendering 
them more exploitable. For access complexity, medium access complexity is a 
majority for both categories, and the difference between them is not large enough to 
significantly affect the score.



\subsection{Vulnerability Response}
\label{sec:vul-resp}


We evaluate the responses to discovered side-channel vulnerabilities from the
cryptographic library developers.

\bheading{Response speed.}
For each vulnerability, we measure the \emph{vulnerability window}, defined as the duration 
from the vulnerability publication date\footnote{A side-channel vulnerability may be published
in different ways, including online archives, security conferences and journals, and the
CVE system. We use the earliest of all such dates.} to the patch
release date. If the patch release date is earlier than the 
vulnerability publication date, the vulnerability window is negative. Obviously narrower
vulnerability window leads to fewer chances of exploit and less damage.

\begin{figure}[t]
  \centering
  \begin{subfigure}{0.5\linewidth}
    \centering
    \includegraphics[width=\linewidth]{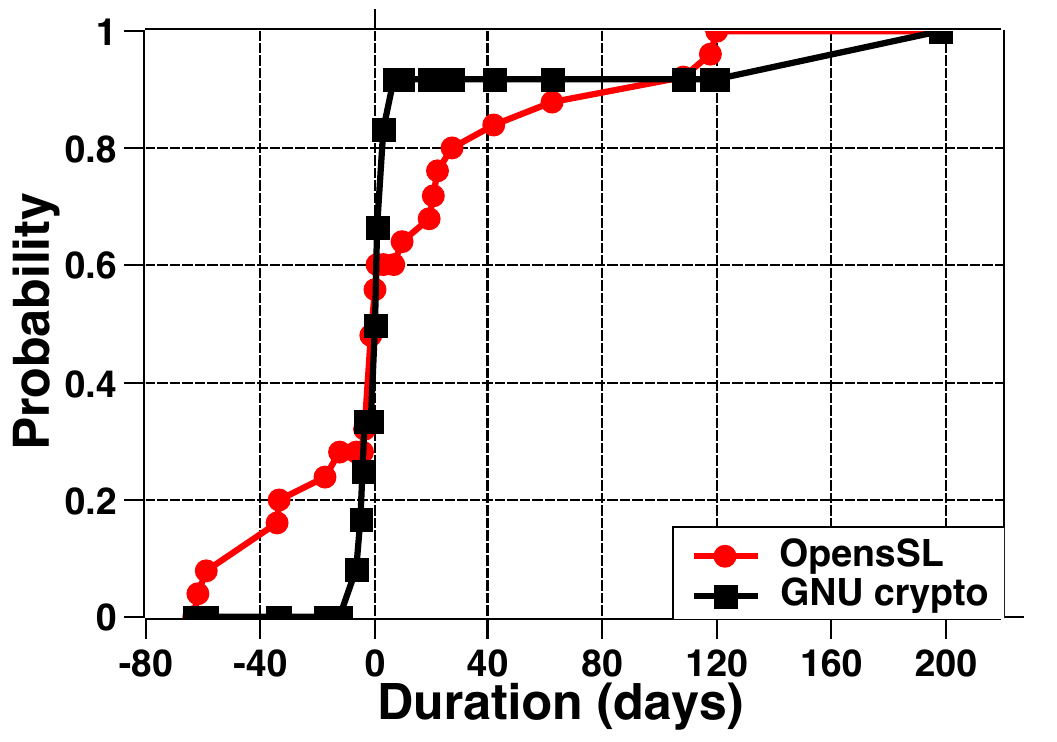}
    \caption{Different libraries}
    \label{fig:cve-3}
  \end{subfigure}%
  \begin{subfigure}{0.5\linewidth}
    \centering
    \includegraphics[width=\linewidth]{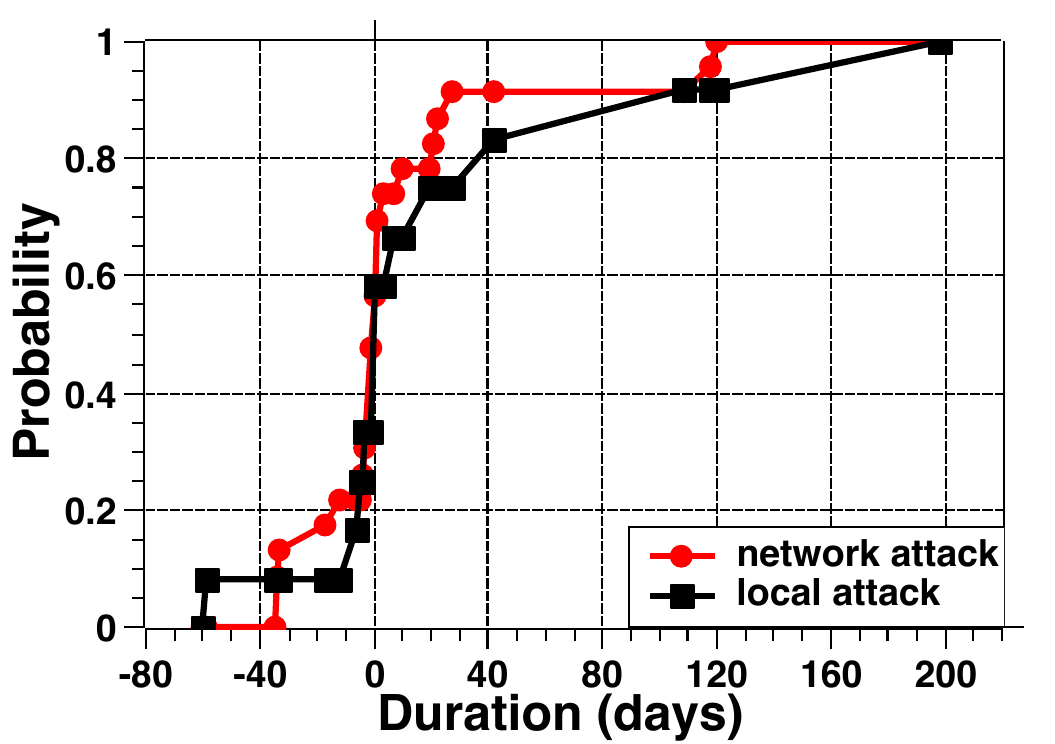}
    \caption{Different access vectors}
    \label{fig:cve-6}
  \end{subfigure}
  \caption{Cumulative distributions of vulnerability windows}
  \label{fig:cumu-dist}
  \vspace{-10pt}
\end{figure}

Figure \ref{fig:cve-3} shows the cumulative distribution of vulnerability windows for OpenSSL
and GNU Crypto. We can see that \textbf{\emph{both libraries responded to side-channel vulnerabilities 
very actively}}: 56\% and 50\% of vulnerabilities were fixed by the two libraries respectively before 
publication; more than 80\% of vulnerabilities were fixed within one month of their 
disclosure; each library has only one case that spanned more than 4 months, the longest 
duration being 198 days in GnuPG.
Figure \ref{fig:cve-6} compares the vulnerability windows of different access vectors. 
\textbf{\emph{Although network attacks are more severe than local attacks, they were fixed at similar
speeds.}}

\bheading{Response coverage.}
We found that \textbf{\emph{most discovered vulnerabilities were addressed in OpenSSL and GNU Crypto, except that
host-based padding oracle vulnerabilities \cite{XiLiCh:17, RoPaSh:18} still exist in both
libraries at the time of writing}}. One possible reason is that such host attacks require stronger adversarial
capabilities and can only work in limited contexts, and thus are less severe. 


\subsection{Cross-branch Patch Consistency}
\label{sec:cross-branch}


An application usually maintains different development branches concurrently. When a 
vulnerability is discovered, if the corresponding patch is not applied to all live
branches at the same time, then an adversary gets an chance to attack the unpatched
branches. For instance, OpenSSL replaced the vulnerable sliding window scalar multiplication with
branchless Montgomery ladder in version 1.1.0i on August 14, 2018, but not in the 1.0.2 branch.
This left a chance for port-based attacks \cite{AlBrHa:19} to work on the sliding window
implementation in OpenSSL 1.0.2, which urged the developers to apply the patch to 1.0.2q on
November 20, 2018.

\begin{figure}[t]
  \centering
  \begin{subfigure}{0.5\linewidth}
    \centering
    \includegraphics[width=\linewidth]{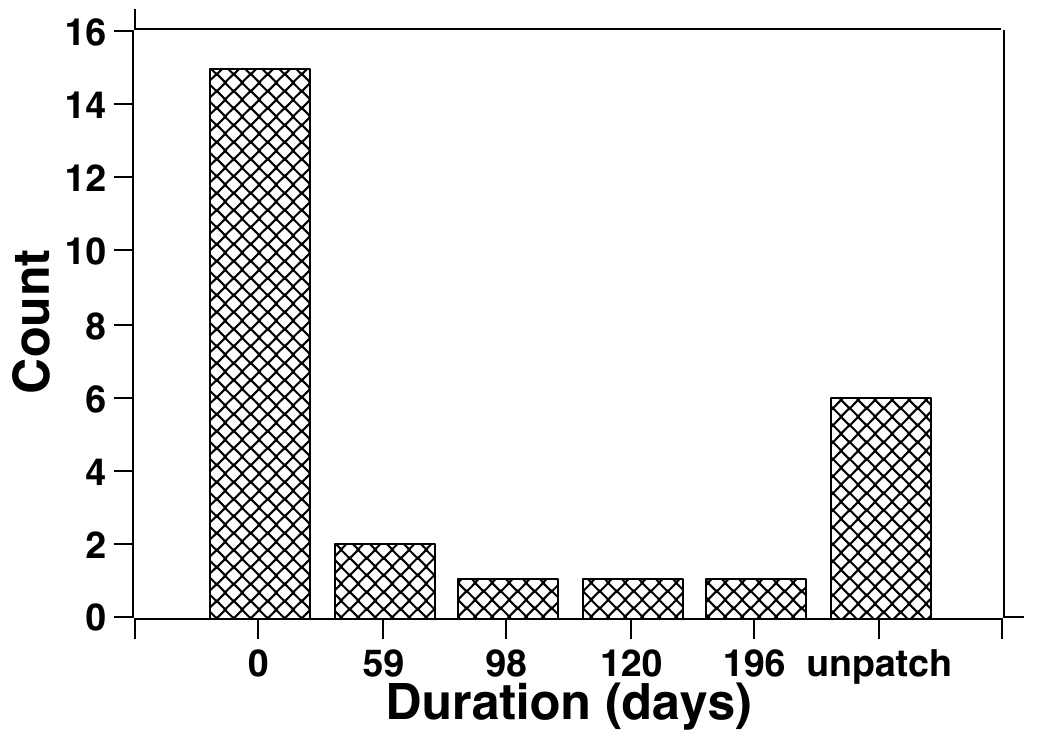}
    \caption{OpenSSL}
    \label{fig:cve-7}
  \end{subfigure}%
  \begin{subfigure}{0.5\linewidth}
    \centering
    \includegraphics[width=\linewidth]{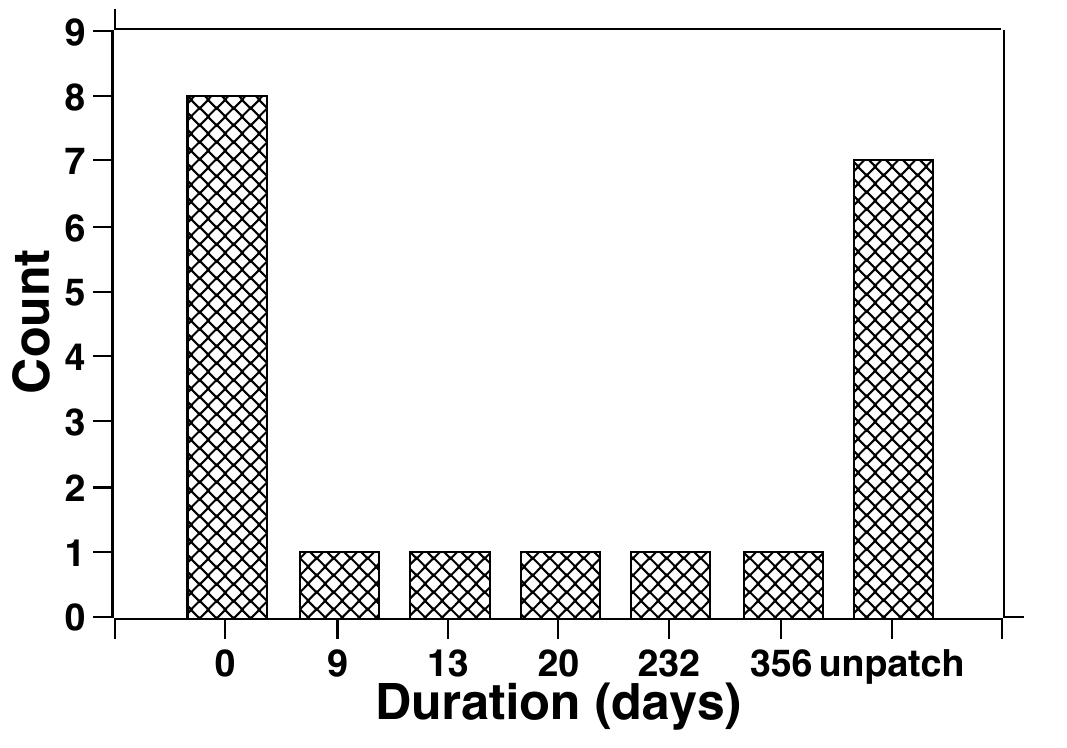}
    \caption{GNU crypto}
    \label{fig:cve-8}
  \end{subfigure}
  \caption{Number of patches for cross-branch windows}
  \label{fig:prob-dist}
  \vspace{-10pt}
\end{figure}

For each vulnerability, we measure the \emph{cross-branch vulnerability window}, defined as
the duration from the first patch release date to the date when all live branches are patched.
Figure \ref{fig:prob-dist} shows the number of patches in different vulnerability windows
for both libraries.
\textbf{\emph{In most cases, a patch was applied to all live branches at the same time (0 days).
Some patches are however still missing in certain branches at the time of writing (unpatch).}}
For example, OpenSSL 1.0.1 introduced masked-window multiplication
and AES-NI support that were never ported to 0.9.8 and 1.0.0 branches before
their end of life. OpenSSL 1.0.2r includes a bug fix for protocol error handling, but it
is not applied to 1.1.0 and 1.1.1. Some new side-channel bug fixes,
not critical though, in OpenSSL 1.1.1 and 1.1.1b are not included in 1.0.2 and 1.1.0. For GNU Crypto,
CVE-2015-0837 was fixed in GnuPG 1.4.19 and Libgcrypt 1.6.3, but not in Libgcrypt 1.5.x. Fortunately
this branch has reached its end of life on December 31, 2016.

\subsection{Countermeasure Type}
\label{sec:count-type}

\begin{figure}[t]
  \centering
  \begin{subfigure}{0.5\linewidth}
    \centering
    \includegraphics[width=\linewidth]{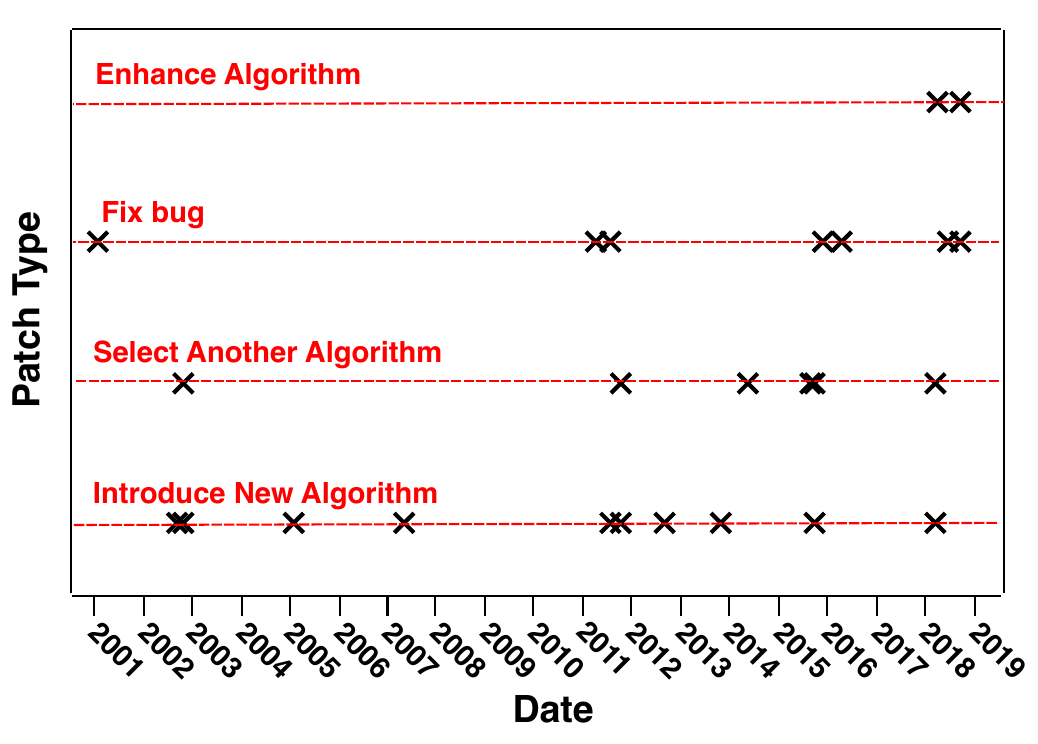}
    \caption{OpenSSL}
    \label{fig:cve-9}
  \end{subfigure}%
  \begin{subfigure}{0.5\linewidth}
    \centering
    \includegraphics[width=\linewidth]{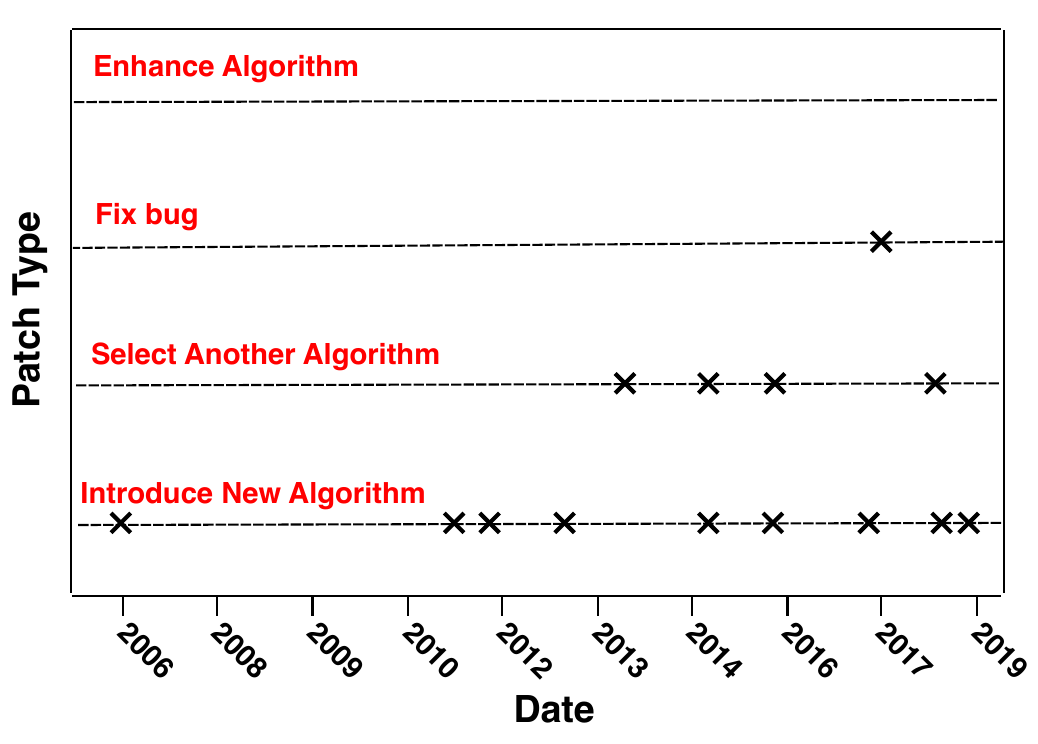}
    \caption{GNU crypto}
    \label{fig:cve-10}
  \end{subfigure}
  \caption{Countermeasure types of the two libraries}
  \label{fig:fix-type}
\end{figure}

We study the types of countermeasures commonly adopted by cryptographic libraries to fix 
side-channel vulnerabilities. Four categories are considered: (1) introducing
brand new implementations; (2) selecting existing secure implementations; (3) fixing software bugs;
(4) enhancing robustness of existing implementations. Classification of countermeasures for
OpenSSL and GNU Crypto is shown in Figure \ref{fig:fix-type}.

In the earlier days, OpenSSL mainly introduced new implementations to fix side-channel 
vulnerabilities. \textbf{\emph{After many years' evolution, each cryptographic operation has secure 
implementations, and brand new solutions become unnecessary.}} Recent patches
were often minor bug fixes. Besides, previously developers only patched the 
code upon revelation of new issues. Now they proactively fortify the
library without the evidence of potential vulnerabilities. This definitely improves the
security of the library against side-channel attacks. 

GNU Crypto has fewer vulnerabilities and patches compared to OpenSSL, and prefers using
traditional solutions for some common issues. For instance, to 
mitigate the vulnerability in sliding window scalar multiplication, OpenSSL
adopted a new solution, masked-window multiplication, while Libgcrypt regressed to
less efficient Double-and-Add-always. Besides, development of GNU Crypto is generally
several years behind OpenSSL. 

\subsection{Comparisons with Other Libraries}
\label{sec:cross-library-comp}

Finally we summarize side-channel CVEs in other cryptographic applications (Table \ref{table:cve-other}), and 
compare them with OpenSSL and GNU Crypto. 

\begin{table}[h]
\caption{Vulnerabilities in other applications}
\label{table:cve-other}
\begin{subtable}[h]{\linewidth}
\centering
\resizebox{0.99\linewidth}{!}{
\begin{tabular}{|l|l|l|l|l|l|}
  \hline
\multicolumn{1}{|l|}{\textbf{Date}}  & \multicolumn{1}{|l|}{\textbf{Application}} & \multicolumn{1}{|l|}{\textbf{Vulnerable Operations}} & \multicolumn{1}{|l|}{\textbf{CVE}} & \multicolumn{1}{|l|}{\textbf{Patch date}} & \multicolumn{1}{|l|}{\textbf{Index}} \\ \hline
2001/06/27  & OpenSSH, AppGate, ssh-1 & RSA-PAD & CVE-2001-0361 & 2001/01/29 & \\ \hline
2004/12/31  & MatrixSSL & Modular Multiplication  & CVE-2004-2682 & 2004/06/01 & \\ \hline
2009/08/31  & XySSL & RSA-PAD & CVE-2008-7128 & ${}^\star$ & \\ \hline
2010/10/20  & Apache MyFaces & Padding oracle attack & CVE-2010-2057 & 2010/06/10 & \\ \hline
2010/10/20  & Oracle Mojarra & Padding oracle attack & CVE-2010-4007 & 2010/06/10 & \\ \hline
2013/02/08  & Rack  & HMAC comparison  & CVE-2013-0263 & 2013/02/07 & \\ \hline
2013/02/08  & Mozilla NSS & MAC-CBC-PAD & CVE-2013-1620 & 2013/02/14 & 4.2.2.b \\ \hline
2013/02/08  & wolfSSL CyaSSL  & MAC-CBC-PAD & CVE-2013-1623 & 2013/02/05 & 4.2.2.b\\ \hline
2013/02/08  & Bouncy Castle & MAC-CBC-PAD & CVE-2013-1624 & 2013/02/10 & 4.2.2.b \\ \hline
2013/10/04  & PolarSSL  & RSA-CRT & CVE-2013-5915 & 2013/10/01 & \\ \hline
2013/11/17  & OpenVPN & Padding oracle attack & CVE-2013-2061 & 2013/03/19 & \\ \hline
2015/07/01  & Libcrypt++  & Rabin-Williams DSA  & CVE-2015-2141 & 2015/11/20 & \\ \hline
2016/04/07  & Erlang/OTP  & MAC-CBC-PAD & CVE-2015-2774 & 2015/03/26 & 4.2.1.b \\ \hline
2016/05/13  & Botan & MAC-CBC-PAD & CVE-2015-7827 & 2015/10/26 & \\ \hline
2016/05/13  & Botan & Modular inversion & CVE-2016-2849 & 2016/04/28 & \\ \hline
2016/07/26  & Cavium SDK  & RSA-CRT & CVE-2015-5738 & ${}^{\star\star}$ & \\ \hline
2016/09/03  & jose-php  & HMAC comparison  & CVE-2016-5429 & 2016/08/30 & \\ \hline
2016/10/10  & Intel IPP & RSA & CVE-2016-8100 & ${}^{\star\star}$ & \\ \hline
2016/10/28  & Botan & RSA-PAD & CVE-2016-8871 & 2016/10/26 & \\ \hline
2016/12/13  & wolfSSL & AES T-table lookup  & CVE-2016-7440 & 2016/09/26 & \\ \hline
2017/01/23  & Malcolm Fell jwt  & Hash comparison & CVE-2016-7037 & 2016/09/05 & \\ \hline
2017/02/13  & Crypto++  &   & CVE-2016-3995 & 2016/09/11 & \\ \hline
2017/03/03  & MatrixSSL & RSA-CRT & CVE-2016-6882 & 2016/11/25 & \\ \hline
2017/03/03  & MatrixSSL & RSA-PAD & CVE-2016-6883 & 2016/04/18 & \\ \hline
2017/03/07  & Intel QAT & RSA-CRT & CVE-2017-5681 & ${}^{\star\star}$ & \\ \hline
2017/04/10  & Botan & MAC-CBC-PAD & CVE-2015-7824 & 2015/10/26 & \\ \hline
2017/04/14  & Nettle  & Modular exponentiation  & CVE-2016-6489 & 2016/08/04 & 1.2.3.e \\ \hline
2017/07/27  & Apache HTTP & Padding oracle attack & CVE-2016-0736 & 2016/10/20 & \\ \hline
2017/08/10  & Apache CXF  & MAC comparison  & CVE-2017-3156 & ${}^{\star\star}$ & \\ \hline
2017/08/20  & Nimbus JOSE+JWT & Padding oracle attack & CVE-2017-12973  & 2017/06/02 & \\ \hline
2017/09/25  & Botan & Modular exponentiation  & CVE-2017-15533  & 2017/10/02 & \\ \hline
2017/12/12  & Erlang/OTP  & RSA-PAD & CVE-2017-1000385  & 2017/11/23 & \\ \hline
2017/12/12  & Bouncy Castle & RSA-PAD & CVE-2017-13098  & 2017/12/28  & 4.1.1.c \\ \hline
2017/12/12  & wolfSSL & RSA-PAD & CVE-2017-13099  & 2017/10/31 & 4.1.1.c \\ \hline
2018/01/02  & Linaro OP-TEE & RSA Montgomery  & CVE-2017-1000413  & 2017/07/07 & \\ \hline
2018/06/04  & Bouncy Castle & DSA & CVE-2016-1000341  & 2016/12/23 & \\ \hline
2018/06/04  & Bouncy Castle & Padding oracle attack & CVE-2016-1000345  & 2016/12/23 & \\ \hline
2018/06/14  & LibreSSL  & Modulo primitive  & CVE-2018-12434  & 2018/06/13 &  \\ \hline
2018/06/14  & Botan & Modulo primitive  & CVE-2018-12435  & 2018/07/02 & \\ \hline
2018/06/14  & wolfssl & Modulo primitive  & CVE-2018-12436  & 2018/05/27 & \\ \hline
2018/06/14  & LibTomCrypt & Modulo primitive  & CVE-2018-12437  & ${}^\star$ & \\ \hline
2018/06/14  & LibSunEC  & Modulo primitive  & CVE-2018-12438  & ${}^\star$ & \\ \hline
2018/06/14  & MatrixSSL & Modulo primitive  & CVE-2018-12439  & 2018/09/13 & \\ \hline
2018/06/14  & BoringSSL & Modulo primitive  & CVE-2018-12440  & ${}^\star$ & \\ \hline
2018/07/28  & ARM mbed TLS  & MAC-CBC-PAD & CVE-2018-0497 & 2018/07/24 & \\ \hline
2018/07/28  & ARM mbed TLS  & MAC-CBC-PAD & CVE-2018-0498 & 2018/07/24 & \\ \hline
2018/09/21  & Apache Mesos  & HMAC comparison & CVE-2018-8023 & 2018/07/25 & \\ \hline
2018/12/03  & nettle  & RSA-PAD & CVE-2018-16869  & ${}^\star$ & 4.1.2.d \\ \hline
2018/12/03  & wolfSSL & RSA-PAD & CVE-2018-16870  & 2018/12/27  & 4.1.2.d \\ \hline
2019/03/08  & Botan & Scalar multiplication & CVE-2018-20187  & 2018/10/01 & \\ \hline
  \end{tabular}
}
\caption{Open-source libraries}
\label{table:cve-opensource}
\end{subtable}

\begin{subtable}[h]{\linewidth}
\centering
\resizebox{0.99\linewidth}{!}{
\begin{tabular}{|l|l|l|l|l|l|}
  \hline
\multicolumn{1}{|l|}{\textbf{Date}}  & \multicolumn{1}{|l|}{\textbf{Application}} & \multicolumn{1}{|l|}{\textbf{Vulnerable Operations}} & \multicolumn{1}{|l|}{\textbf{CVE}} & \multicolumn{1}{|l|}{\textbf{Patch date}} & \multicolumn{1}{|l|}{\textbf{Index}} \\ \hline
2010/09/22  & Microsoft IIS & Padding oracle attack & CVE-2010-3332 & 2010/09/27 & \\ \hline
2013/02/08  & Opera & MAC-CBC-PAD & CVE-2013-1618 & 2013/02/13 & 4.2.2.b \\ \hline
2013/06/21  & IBM WebSphere Commerce & Padding oracle attack & CVE-2013-0523 & ${}^{\star\star}$ & \\ \hline
2014/08/16  & IBM WebSphere DataPower & - & CVE-2014-0852 & ${}^{\star\star}$ & \\ \hline
2014/12/09  & F5 BIG-IP & MAC-CBC-PAD & CVE-2014-8730 & ${}^{\star\star}$ & \\ \hline
2015/08/02  & Siemens RuggedCom ROS & MAC-CBC-PAD & CVE-2015-5537 & ${}^{\star\star}$ & \\ \hline
2015/11/08  & IBM DataPower Gateways  & Padding oracle attack & CVE-2015-7412 & ${}^{\star\star}$ & \\ \hline
2016/04/12  & EMC RSA BSAFE & RSA-CRT & CVE-2016-0887 & ${}^{\star\star}$ & \\ \hline
2016/04/21  & CloudForms Management Engine  & Padding oracle attack & CVE-2016-3702 & ${}^\star$ & \\ \hline
2016/09/08  & HPE Integrated Lights-Out 3 & Padding oracle attack & CVE-2016-4379 & 2016/08/30  & 4.2.1.a \\ \hline
2016/12/15  & Open-Xchange OX Guard  & Padding oracle attack & CVE-2016-4028 & 2016/04/21 & \\ \hline
2017/02/03  & EMC RSA BSAFE & Padding oracle attack & CVE-2016-8217 & 2017/01/20 & \\ \hline
2017/03/23  & Cloudera Navigator  & MAC-CBC-PAD & CVE-2015-4078 & ${}^{\star\star}$ & \\ \hline
2017/06/30  & OSCI-Transport  & Padding oracle attack & CVE-2017-10668  & ${}^\star$ & \\ \hline
2017/08/02  & Citrix NetScaler  & MAC-CBC-PAD & CVE-2015-3642 & ${}^{\star\star}$ & \\ \hline
2017/11/17  & F5 BIG-IP  & RSA-PAD & CVE-2017-6168 & ${}^{\star\star}$ & 4.1.1.c \\ \hline
2017/12/13  & Citrix NetScaler  & RSA-PAD & CVE-2017-17382  & ${}^{\star\star}$ & 4.1.1.c \\ \hline
2017/12/13  & Radware Alteon  & RSA-PAD & CVE-2017-17427  & ${}^{\star\star}$ & 4.1.1.c \\ \hline
2017/12/15  & Cisco ASA & RSA-PAD & CVE-2017-12373  & 2018/01/05 & 4.1.1.c \\ \hline
2018/01/10  & Palo Alto Networks PAN-OS & RSA-PAD & CVE-2017-17841  & ${}^{\star\star}$ & 4.1.1.c \\ \hline
2018/02/05  & Cavium Nitrox and TurboSSL  & RSA-PAD & CVE-2017-17428  & ${}^{\star\star}$ & 4.1.1.c \\ \hline
2018/02/07  & IBM GSKit & Padding oracle attack & CVE-2018-1388 & ${}^{\star\star}$ & \\ \hline
2018/02/26  & Unisys ClearPath MCP  & RSA-PAD & CVE-2018-5762 & ${}^{\star\star}$ & 4.1.1.c \\ \hline
2018/05/17  & Symantec SSL Visibility & RSA-PAD & CVE-2017-15533  & 2018/01/12 & 4.1.1.c \\ \hline
2018/05/17  & Symantec IntelligenceCenter & RSA-PAD & CVE-2017-18268  & ${}^{\star\star}$ & 4.1.1.c \\ \hline
2018/07/31  & Huawei products & RSA-PAD & CVE-2017-17174  & ${}^{\star\star}$ & \\ \hline
2018/08/15  & Clavister cOS Core  & RSA-PAD & CVE-2018-8753 & ${}^{\star\star}$ & \\ \hline
2018/08/15  & ZyXEL ZyWALL/USG  & RSA-PAD & CVE-2018-9129 & ${}^{\star\star}$ & \\ \hline
2018/08/21  & Huawei products & RSA-PAD & CVE-2017-17305  & ${}^{\star\star}$ & \\ \hline
2018/08/31  & RSA BSAFE Micro Edition Suite & RSA-PAD & CVE-2018-11057  & ${}^{\star\star}$ & \\ \hline
2018/09/11  & RSA BSAFE SSL-J & RSA-PAD & CVE-2018-11069  & ${}^{\star\star}$ & \\ \hline
2018/09/11  & RSA BSAFE Crypto-J  & RSA-PAD & CVE-2018-11070  & ${}^{\star\star}$ & \\ \hline
2018/09/12  & Intel AMT  & RSA-PAD & CVE-2018-3616 & ${}^{\star\star}$ & \\ \hline
2019/02/22  & Citrix NetScaler Gateway  & Padding oracle attack & CVE-2019-6485 & ${}^{\star\star}$ & \\ \hline
  \end{tabular}
}
  \caption{Closed-source products}
  \label{table:cve-closesource}
  \end{subtable}
\begin{tablenotes}
  \item ${}^\star$ Whether and when this vulnerability is addressed is not revealed.
  \item ${}^{\star\star}$ This vulnerability is addressed, but the date is not revealed.
\end{tablenotes}
\end{table}

\begin{figure}[t]
  \centering
  \begin{subfigure}{0.49\linewidth}
    \centering
    \includegraphics[width=\linewidth]{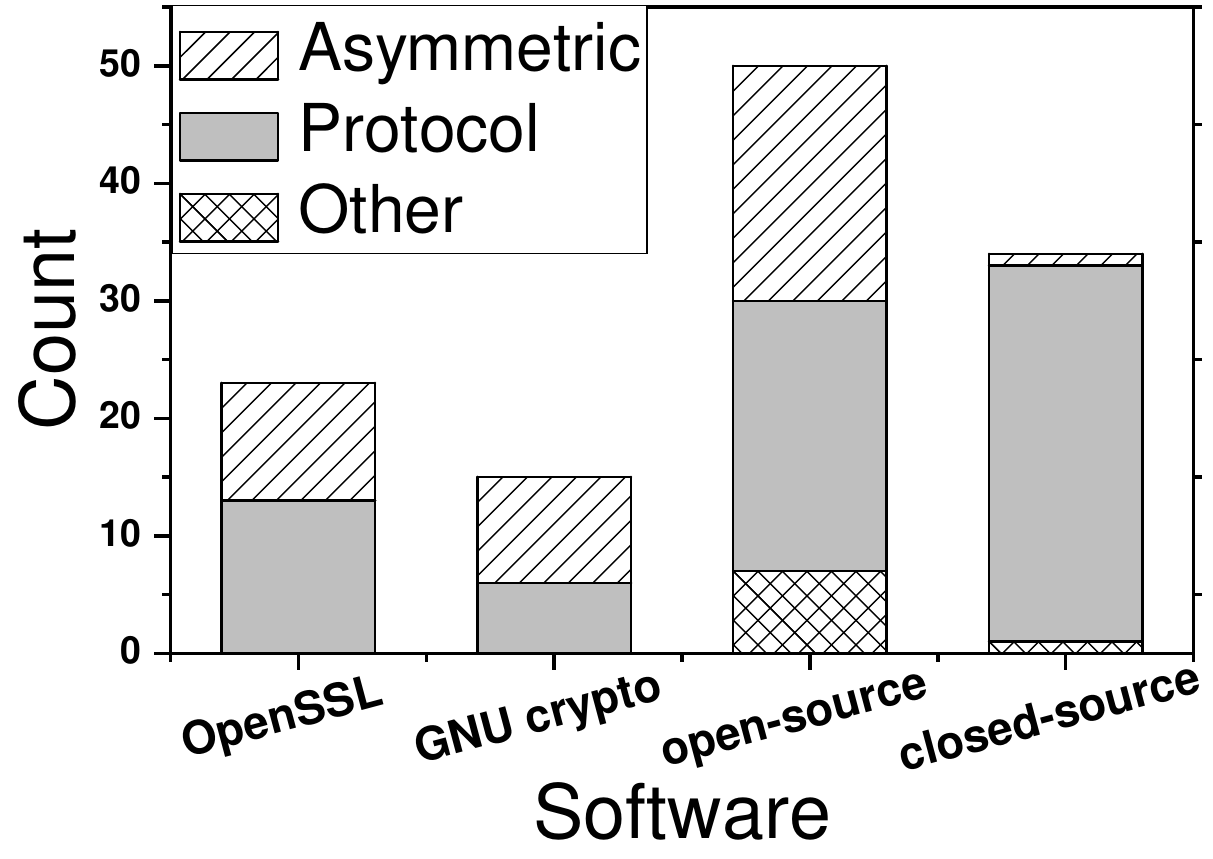}
    \caption{Vulnerable categories}
    \label{fig:cve-11}
  \end{subfigure}%
  \begin{subfigure}{0.51\linewidth}
    \centering
    \includegraphics[width=\linewidth]{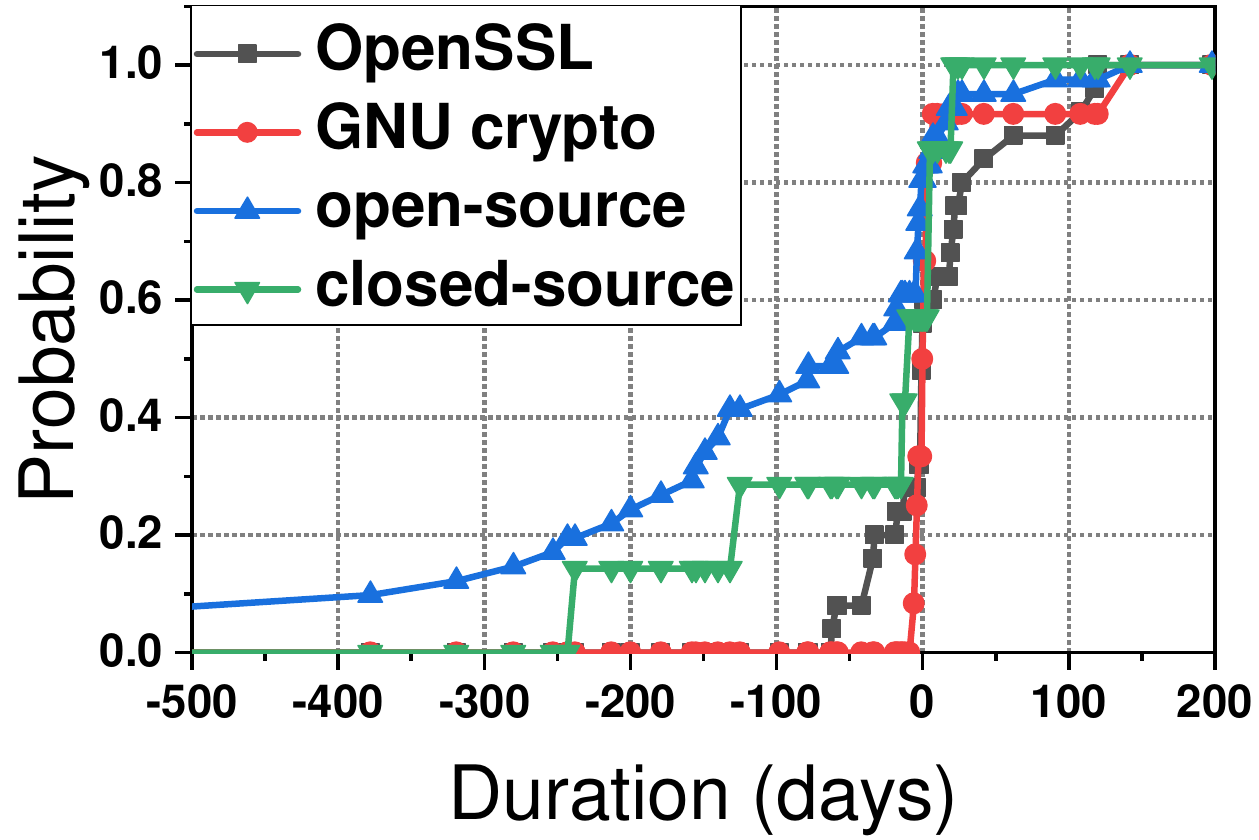}
    \caption{Response speed}
    \label{fig:cve-12}
  \end{subfigure}
  \caption{Comparisons between different applications}
  \label{fig:other-lib}
\vspace{-10pt}
\end{figure}

\bheading{Vulnerable Categories.}
Figure \ref{fig:cve-11} shows the breakdown of vulnerabilities in different categories.
We observe that \textbf{\emph{side-channel vulnerabilities exist widely in many applications, in 
addition to OpenSSL and GNU Crypto}}. We believe a lot of 
unrevealed side-channel vulnerabilities still exist in various applications, for two reasons.

First, researchers tend to study common cryptographic libraries, encouraging their developers
to continuously improve the code. Other less evaluated 
applications may still contain out-of-date vulnerabilities, but their developers are unaware or
ignorant of them. For instance, Bleichenbacher attack was proposed 20 
years ago and has been mitigated in common libraries like OpenSSL and GnuTLS, but it
still exists in about one third of top 100 Internet domains including
Facebook and PayPal, as well as widely used products from IBM, Cisco and so on \cite{HaJuCr:18}. 

Second, some attacks such as those against cryptographic primitives or based on 
micro-architectural contention require the source code to be available, prohibiting
researchers from discovering vulnerabilities in closed-source applications. 
For instance, Figure \ref{fig:cve-11} shows that the majority of vulnerabilities found in 
closed-source applications are padding oracles via remote timing or message side channels,
likely because no source code is needed to experiment with these attacks.
We do not know if they also suffer from padding oracle attacks via micro-architectural side
channels, as current studies \cite{IrLnEi:15, XiLiCh:17, RoPaSh:18, RoGiGe:19} evaluated
them only on open-source libraries. It is also unclear if they possess vulnerabilities
related to modular or scalar operations for the similar reason.


\bheading{Response speed and coverage.}
Figure \ref{fig:cve-12} compares the response speeds of different applications.
Interestingly, \textbf{\emph{they all responded to the vulnerabilities very fast.}} Most vulnerabilities
were published only after the release of corresponding patches, leaving 
no vulnerability windows for adversaries to exploit.

Regarding the coverage, \textbf{\emph{most discovered vulnerabilities were addressed, with a few exceptions}}
(annotated with ${}^\star$ in Tables \ref{table:cve-opensource} and \ref{table:cve-closesource})
where too little public information is available. For these cases, we are unable to ascertain
whether these issues were solved or not.

\section{Insights and Lessons}
\label{sec:lesson}

We draw some insights and suggestions based on the summarization and 
evaluation of side-channel vulnerabilities and countermeasures. We hope they can benefit 
cryptographic software developers, users and researchers.

\subsection{Research Directions}
First, we propose several promising opportunities and directions for future side-channel research. 

\bheading{D1: Identify new vulnerabilities in state-of-the-art cryptographic implementations, more likely in less-studied ciphers and operations.}


The discovery and exploitation of various side-channel vulnerabilities during the past decades
inspired people to search for new secure solutions. Now each critical cryptographic primitive
has secure algorithms and implementations that are robust against different types of side-channel attacks. 
Thus, it is interesting
to explore new fundamental vulnerabilities in current implementations. In addition to the cryptographic
operations discussed in \secref{sec:crypto}, researchers can focus on other ciphers and operations, and
search for new critical vulnerabilities.
For instance, the emerging post-quantum cryptography has drawn people's attention for its resistance against quantum computer based attacks. Different types of
post-quantum algorithms were designed for public key infrastructure, e.g., hash-based, lattice-based and multivariate cryptography. Currently researchers only focus on the
side-channel vulnerabilities in lattice-based cryptography. It is interesting to explore
if there are any chances to attack other types of post-quantum cryptography schemes. 

\bheading{D2: Discover outdated vulnerabilities in less popular and closed-source
applications.}

As discussed in \secref{sec:cross-library-comp}, old vulnerabilities, especially those against cryptographic primitives or 
via micro-architectural contention, can still exist in real-world applications. It is worth the research effort to conduct a more
comprehensive evaluation of various side-channel vulnerabilities on different open-source 
libraries and commercial products. This will raise public awareness on side-channel threats, 
and more importantly, encourage developers to secure these applications as a result.

\bheading{D3: Design new side-channel attack techniques that can capture finer grained side-channel information, and study countermeasures for the new attacks.}

From \secref{sec:crypto} we observe that innovation in providing adversaries with more detailed information
usually exposes new vulnerabilities. For instance, fixed window modular exponentiation with 
scatter-gather approach was designed to defeat attacks with 
the granularity of cache line. However, a new attack method that 
captures cache-bank level information \cite{YaGeHe:17} invalidated the approach. Previous 
countermeasures for padding oracle attacks only considered remote attacks, and thus they
focused on producing uniform responses in constant time. Later, host-based attack techniques
were applied to padding oracle settings \cite{RoGiGe:19, IrLnEi:15, XiLiCh:17, RoPaSh:18},
compromising such solutions. We hence propose this research direction so
the security of cryptographic operations can be constantly enhanced.

\subsection{Development Recommendations}
We offer some suggestions for developers of cryptographic applications and systems to make
the implementations more secure and efficient.

\bheading{R1: Different countermeasure strategies can have different vulnerability coverages
and performance costs. Developers need to carefully select the optimal strategies for critical
operations.}

We summarized different countermeasures and their features in \secref{sec:insights}. 
To remove control flow vulnerabilities, AlwaysExecute strategies introduce performance 
overhead due to execution of all possible code routines. ConditionalSelect can only defeat 
remote attacks since it still leaves a minor secret-dependent control flow, while 
BitwiseSelect is secure against all known types of attacks.

To remove data flow vulnerabilities, AlwaysAccess-BitwiseSelect defeats all known types of 
attacks with the price of performance degradation from extra memory accesses. On-the-fly 
calculation is useful if the computation does not contain secret-dependent control flows, while repeated computation causes slowdown.

Cryptographic blinding is a technique dedicated for asymmetric cryptography such as RSA, ElGamal and ECC.
It is particularly effective in cases where secret-dependent control flow or data flow 
is hard to remove (e.g., modular multiplication).
Plaintext/ciphertext blinding can only defeat chosen-input side-channel attacks, while key 
blinding does not have this limitation.

\bheading{R2: Software bugs contribute a lot to side-channel vulnerabilities.
Developers should verify the code correctness of critical operations prior to release.}

Developers may accidentally leave software bugs that can be exploited to perform
side-channel attacks. As critical cryptographic operations are designed to be more
secure today, adversaries now tend to search such bugs to break the secure
designs and implementations (\secref{sec:count-type}). There are three common types of 
side-channel bugs. 
First, incorrect flag propagation prevents some cryptographic
operations from calling the securely implemented routine even though it is already introduced,
exemplified by modular exponentiation \cite{GaBrYa:16} and inversion
\cite{GaPeBr:16}. Second, some secure implementations are only partly applied to the
operations or primitives, leaving the rest still vulnerable. 
For instance, in OpenSSL before version 1.0.2q, most curves (P-224, P-256 and P-521) adopted
secure branchless Montgomery ladder or masked window for scalar multiplication,
but P-384 curve still used the insecure sliding window implementation, offering the adversary
an opportunity to compromise ECDSA \cite{AlBrHa:19}. Third, some corner cases were overlooked
during implementation, giving rise to non-constant runtime behaviors in execution time
\cite{AlPa:16, RoPaSh:18}, message type \cite{MeSoJu:14} and format \cite{AvScSo:16}.

To verify if the secure routines are selected properly and applied to all operations or 
primitives, developers can insert
breakpoints inside the insecure routines and check if they are reached \cite{AlGaTa:18}.
To verify if the implementation has expected constant behaviors, developers can measure
the side-channel information under different cases and check if they are indistinguishable.

\bheading{R3: Developers should carefully handle the reported vulnerabilities. They should (1) prioritize the most severe vulnerabilities; (2) apply patches to all live
branches at the same time; (3) apply patches to all vulnerable operations.}

From \secref{sec:vul-severe} we observe that different types of side-channel vulnerabilities
are assigned with different severity scores. The network-level vulnerabilities are 
particularly severe as they are more exploitable, so developers should pay special attention 
to those attacks and prioritize their fixes. Besides, when a vulnerability is 
disclosed, developers should apply the corresponding countermeasure patches to all live 
branches of the application immediately. Otherwise, the unpatched branch will remain 
vulnerable as the threat is disclosed to the public (\secref{sec:cross-branch}). Developers 
should also check whether the application contains other implementations sharing the 
same vulnerability. If so they should also fix them even though they are not reported. 

\subsection{Indications in Practice}
We draw some conclusions for cryptographic application users about side-channel threats and
countermeasures from a practical viewpoint.

\bheading{I1: Side-channel vulnerabilities are generally considered less severe compared to
other vulnerabilities, although their volume in CVE database is non-trivial.}

As discussed in \secref{sec:vul-severe}, two reasons make side-channel threats less severe:
the attacks require higher adversarial capabilities and more domain knowledge,
and the damage is relatively smaller. Thus, the existence of these vulnerabilities in 
applications is by no means a catastrophe to the users and there is no need to panic.
The world has not seen an occurrence of malicious side-channel attacks that could bring severe damages so far.

\bheading{I2: Although side-channel vulnerabilities are regarded as less severe, they are handled
very actively, either in mainstream open-source libraries (\secref{sec:vul-resp}), less
popular libraries or closed-source products (\secref{sec:cross-library-comp}).}

Such active attitude towards side-channel threats is reflected in several aspects. First, new
side-channel vulnerabilities were fixed at a very fast speed to reduce users' potential damage.
Second, most known side-channel vulnerabilities to date were fixed. Third, cryptographic libraries
are now shifting the countermeasures from brand new implementations to minor bug fixes and
enhancement, indicating that they are becoming more mature (\secref{sec:count-type}).

\bheading{I3: It is unnecessary for users to always patch or upgrade the libraries for the most
secure implementations. They can adopt those suited to their scenarios.}

If a use case does not need to consider certain adversarial capabilities, countermeasures designed
for them will not be necessary. For instance, if the application does not have network interfaces,
then the user does not need to consider network-level side-channels. If the application does not
share the computing platform with other untrusted parties, then the user does not need to consider
host-level attacks.

\section{Related Work}
\label{sec:relate}

\subsection{Physical Attacks}
Different from software side-channel attacks, physical side-channel attacks require the adversary to
be local to the target system running the victim application to collect the 
side-channel signals during the application execution. Different attacks have been proposed 
against cryptographic applications \cite{Co:99, ArTh:07, GeShTr:14, GePiTr:15, GePaPi:15, 
GePaPi:16, BeFoMa:16, GePaPiIt:16}. Some of them are so powerful that even the application
does not contain secret-dependent control flow or data flow, they are still able to deduce
secrets via intermediate values \cite{Co:99}. Systematization of these attacks and 
vulnerabilities is however not relevant in this paper.

\subsection{Attacking Non-cryptographic Applications}
In addition to recovering keys in cryptographic operations, side-channel attacks can also be used to steal other 
types of information. At the application level, attacks exist to identify keystrokes
\cite{SoWaTi:01, ZhWa:09, ChWaWa:10, lipp2016armageddon, ViKo:17, LiGrSc:17, WaNeQi:19, 
liu2019human}, application states and activities \cite{DiLiLi:16, SpPaMa:18, SpKiGr:18},
and websites \cite{oren2015spy, ShKaHa:18}. At the system level, adversaries may use side 
channels to infer host configurations \cite{schwarz2019javascript} and memory layout information
\cite{HuWiHo:13, EvPoAb:16, JaLeKi:16, GrMaFo:16}. Meltdown \cite{MoMiDa:18} and Spectre
\cite{KoHoFo:19} attacks were disclosed to bypass the isolation and protection schemes
in operating systems, followed by variants of such attacks \cite{JoMaOf:18, ChChXi:19, 
EvRiAb:18, TrLuMa:18, SaAhId:19, ClJoMi:18}. Systematization of these attacks is out of the scope
of our work.

\subsection{Identification of Vulnerabilities}
Various approaches and tools were designed to automatically identify potential
side-channel vulnerabilities in commodity software. Some static methods utilize
abstract interpretation to analyze the source code and measure the information
leakage (bounds) \cite{MoPiSc:05, DoKoMa:15, RoQuAr:16, DoKo:17, BiBuKr:17, FaGuLe:18, WaBaLi:19, BrLiZh:19}, while
others verify if a program always exhibits constant-time behavior by checking
whether it contains secret-dependent control flow or data flow
\cite{BaBeGu:14, AlBaBa:16, BaPiTr:17, DeFrHo:17, BoHaKa:17}. Dynamic methods
profile program execution and measure the correlation between critical secrets
and runtime behaviors from execution traces to capture side-channel vulnerabilities
\cite{ZaHeSi:16, WaWaLi:17, IrCoGu:17, XiLiCh:17, ReBaVe:17, WiMoEi:18, ShKiKw:18, WeZaSp:18}.

\subsection{Side-channel Surveys}
Past efforts summarized side-channel attacks and countermeasures in different
contexts including smart card \cite{Tu:17}, networked systems \cite{ZaArBr:07, BiGhNa:17},
hardware architecture \cite{GeYaCo:18, Sz:18}, cloud \cite{UlZsFa:17, BeWeMu:17},
smartphone \cite{SpMoKo:18, XuSoJi:16, Na:16} and key logging \cite{Mo:18, HuAlZa:16}.
They mostly studied side-channel information collecting techniques for 
specific attack types in specific environments. In contrast, our work focuses on the
vulnerabilities in cryptographic implementations.

Closer to our work are efforts \cite{Av:05, FaGuDe:10, FaVe:12} that studied
side-channel vulnerabilities and countermeasures in Elliptic Curve Cryptosystems (ECC). 
We distinguish our work from theirs by considering a wider range of cryptographic ciphers and
protocols than ECC. We also systematically evaluate
the vulnerabilities and countermeasures in different applications in addition to
reviewing past literature.

\section{Conclusion}
\label{sec:conclu}
Side-channel attacks against cryptographic implementations have been an enduring
topic over the past 20 years. Many vulnerabilities have been discovered from previous cryptographic
implementations, but unknown ones likely still exist in today's implementations.
The good news is that the community resolved these vulnerabilities very actively,
and hence large-scale side-channel attacks causing severe real-world damages
haven't happened so far. Besides, years of efforts have fortified common cryptographic
libraries and applications against side-channel attacks, and recently discovered
vulnerabilities were less significant or surprising.

Looking ahead, we expect continuous arms race between side-channel attacks and defenses.
We encourage researchers to discover new vulnerabilities and attacks,
evaluate them on a wider range of applications, and develop novel countermeasures for them.

\bibliographystyle{IEEEtran}
\bibliography{ref}

\appendix

\subsection{Cryptographic Ciphers and Protocols}
\label{sec:append-crypto}

\subsubsection{Elliptic Curve Cryptography}
\label{sec:append-ecc}
In geometry, an elliptic curve is a two-dimensional curve defined by $y^2 = x^3 + ax + b$.
When used in cryptography, we make a few modifications. We first require $4a^3 + 27b^2 \neq 0$
to exclude singularity. Additionally we apply the curve over a finite field, usually with a prime
or $2^m$ order, and introduce a special infinity point $\infty$. This forms a group of points, 
with $\infty$ as identity element and the $+$ operation defined as: 1) $P+\infty=P$;
2) if $P$ is the reflection of $Q$ over the x-axis, then $P+Q=\infty$; 3) if $P$ and $Q$ are
different points not in case 2, then the line through $P$ and $Q$ intersects the curve
with another point $R$, then $P+Q=R'$ where $R'$ is the reflection of $R$ over the x-axis;
4) if $P$ and $Q$ are the same point, treat ``the line through
$P$ and $Q$'' as ``the line tangent to the curve at $P$'' and use case 3. Case 3 and 4 are non-trivial cases
and can be computed with Equation \ref{eq:pointadd}. We usually write $P+...+P$ as $nP$, and name
this operation ``scalar multiplication''. 


\begin{align}
\label{eq:pointadd}
\footnotesize
\begin{split}
& \lambda = 
\begin{cases}
    \frac{y_q - y_p}{x_q - x_p}, & \text{if } x_q\neq x_p\\
    \frac{3x^2_p + a}{2y_p},              & \text{otherwise}
\end{cases} \\
&x_r = \lambda^2 - x_p - x_q \\
&y_r = \lambda(x_p - x_r) - y_p
\end{split}
\end{align}

Elliptic curve can be used in cryptography \cite{Mi:85, Ko:87}. Assuming the group has
prime order $p$ and generator $G$, choose a random integer $k$ from $[1, p-1]$ as
the private key, then $D=kG$ will be the public key. 

\bheading{Usage in Digital Signature (ECDSA)}. To sign message $m$, Alice chooses a public
hash function $h$ and a secret nonce $n$. Compute $R = nG \mod p$, $r = R_x$ (cannot be zero),
and $s = k^{-1}(h(m)+kr) \mod p$, then $(r,s)$ is the signature. To verify, Bob computes
$u_1 = h(m)s^{-1} \mod p$, $u_2 = rs^{-1} \mod p$, $Z = u_1G + u_2D$, and $z = Z_x$. The
signature is valid if $r \equiv z \mod p$ holds.

\bheading{Usage in Key Exchange (ECDH)}. Assuming Alice uses private key $k_A$ and public
key $P_A=k_AG$, and Bob similarly uses key pair $(k_B, P_B)$, the shared symmetric key is
simply $k = k_BP_A = k_AP_B = (k_Ak_B)G$.

\subsubsection{Key Exchange and Encryption Protocols}
\label{sec:append-pad}

In the handshake protocol, the client first sends a list of cipher suites $cs_c$ and a nonce $r_c$.
The server responds with a list of cipher suites $cs_s$, the server certificate, and a nonce $r_s$.
Then the client picks a cipher $cs$ (e.g., RSA) supported by both the client and server.

The client generates a random secret string $k$ with byte length $l_k$ as the master key.
We require $l_k \leqslant l_n-11$, where $l_n$ is the byte length of $n$ in the server's
public key $(n, e)$. The client builds a message block $m$ (Equation \ref{eq:rsa}a) using a non-zero
random padding string \texttt{pad} with byte length $l_n - 3 – l_k$, which is at least 8.
Then she encrypts $m$ to get the ciphertext (Equation \ref{eq:rsa}b).

\begin{subequations}
\label{eq:rsa}
\footnotesize
  \begin{align}
& m = 0x00 || 0x02 || \texttt{pad} || 0x00 || k \\
& c \equiv m^e \mod n
  \end{align}
\end{subequations}

The server decrypts the received ciphertext and validates if the message format complies with
Equation \ref{eq:rsa}a. If yes, the server sends a finished message to the client,
and the client replies a finished message, marking the completion of the key exchange of $k$.

Equation \ref{eq:mac-cbc} shows how to use CBC-MAC to encrypt message $m$ with MAC
built from a block cipher of block size $b$. (1) The MAC $HM$ is calculated over
the sequence number \texttt{SEQ}, header \texttt{HDR} and message $m$ (Equation \ref{eq:mac-cbc}a).
(2) The plaintext $P$ is created by concatenating $m$, $HM$, and a padding string
\texttt{pad} (Equation \ref{eq:mac-cbc}b) chosen so that the byte length of P
is $Nb$, where $N$ is an integer. The most common way is to pad $n+1$
bytes with each byte value as $n$, e.g., $0x02||0x02||0x02$. (3)
$P$ is divided into a sequence of blocks of $b$ bytes, $p_1, p_2, ..., p_N$, and encrypted
with key $K$ (Equation \ref{eq:mac-cbc}c). (4) The text $T$ transmitted over the network
is the concatenation of \texttt{HDR} and each ciphertext block (Equation \ref{eq:mac-cbc}d).

\begin{subequations}
\label{eq:mac-cbc}
\footnotesize
\vspace{-4pt}
  \begin{align}
& HM = H((K \oplus \texttt{opad}) || H((K \oplus \texttt{ipad})||\texttt{SEQ} || \texttt{HDR} || m)) \\
& P = m || HM || \texttt{pad} \\
& c_i = Enc_K(c_{i-1} \oplus p_i); i = 1, 2, ..., N \\
& T = \texttt{HDR} || c_1 || c_2 || .. || c_N
  \end{align}
\end{subequations}

The receiver decrypts $T$ and accepts the message $m$ only if the padding format and MAC are correct.

\subsection{Implementations of Cryptographic Operations}
\label{sec:append-implement}

We show the pseudo code of implementations for different cryptographic operations described in
Section \ref{sec:crypto}.

\RestyleAlgo{ruled}
\SetAlgoLined
\LinesNumbered

{\centering
\removelatexerror
\begin{minipage}{\linewidth}
\begin{algorithm}[H]
\scriptsize
\SetAlgoLined
 \KwIn{$x$:$x_{k-1}x_{k-2}...x_0$, $y$:$y_{n-1}y_{n-2}...y_0$}
 \Indp 
\Indm \KwOut{$x*y$}
\Indp
\Indm
\SetKwFunction{proc}{MULTIPLY}
\SetKwProg{myalg}{function}{}{end}
    \BlankLine
    \BlankLine
  \myalg{\proc{$x, y$}}{
  \If{\emph{$\texttt{SIZE\_OF\_LIMBS}(x) < \texttt{KARATSUBA\_THRESHOLD}$}} {
    $r\gets \texttt{MUL\_BASECASE}(x, y)$
  }
  $r \gets 0$ \\
  $i \gets 1$ \\
  \While{$i*n \leq k$}{
    $t \gets \texttt{MUL\_KARATSUBA}(x_{i*n-1}...x_{(i-1)*n}, y)$ \\
    $r \gets \texttt{ADD\_WITH\_OFFSET}(r, t, (i-1)*n)$ \\
    $i \gets i+1$
  }
  \If{$i*n>k$}{
    $t \gets \texttt{MULTIPLY}(b, a_{k-1}...a_{(i-1)*n})$ \\
    $r \gets \texttt{ADD\_WITH\_OFFSET}(r, t, (i-1)*n)$ \\
  }
  \KwRet $r$
}

\SetKwFunction{proc}{MUL\_BASECASE}
\SetKwProg{myalg}{function}{}{end}
    \BlankLine
    \BlankLine
\myalg{\proc{$x, y$}}{
  \uIf{$y_0 = 0$}{
    $r \gets 0$
  }\uElseIf{$y_0 = 1$}{
      $r \gets x$
  }\uElse{
      $r \gets \texttt{MUL\_BY\_SINGLE\_LIMB}(x, y_0)$
  }

  \For{$i \gets 1$ \textbf{\emph{to}} $n-1$ } {
    \uIf{$y_i = 1$}{
      $r \gets \texttt{ADD\_WITH\_OFFSET}(r, x, i)$
    }\uElseIf{$y_i > 1$}{
      $r \gets \texttt{MUL\_AND\_ADD\_WITH\_OFFSET}(r, x, y_i, i)$
    }
  }

  \KwRet $r$
}

\SetKwFunction{proc}{MUL\_KARATSUBA}
\SetKwProg{myalg}{function}{}{end}
    \BlankLine
    \BlankLine
\myalg{\proc{$x, y$}}{
  \If{\emph{$n < \texttt{KARATSUBA\_THRESHOLD}$}}{
    \KwRet $\texttt{MUL\_BASECASE}(a, b)$
  }
  \eIf{$n \emph{\text{ mod }} 2 = 1$}{
    $r \gets \texttt{MUL\_KARATSUBA}(x_{n-2}...x_0, y_{n-2}...y_0)$ \\
    $r \gets \texttt{MUL\_AND\_ADD\_WITH\_OFFSET}(r, x_{n-2}...x_0, y_{n-1}, n-1) $ \\
    $r \gets \texttt{MUL\_AND\_ADD\_WITH\_OFFSET}(r, y, x_{n-1}, n-1) $ \\
  }
  {
    $h \gets \texttt{MUL\_KARATSUBA}(x_{n-1}...x_{n/2}, y_{n-1}...y_{n/2})$ \\
    $t \gets \texttt{MUL\_KARATSUBA}(x_{n-1}...x_{n/2}-x_{n/2-1}...x_{0}, y_{n/2-1}...y_{0}-y_{n-1}...y_{n/2})$ \\
    $l \gets \texttt{MUL\_KARATSUBA}(x_{n/2-1}...x_{0}, y_{n/2-1}...y_{0})$ \\
    $r \gets (2^{2*32*n} + 2^{32*n})*h + 2^{32*n}*t + (2^{32*n}+1)*l$
  }
  \KwRet $r$
}
 \caption{Modular Multiplication}
 \label{alg:mul}
\end{algorithm}
\end{minipage}
\par
}

{\centering
\removelatexerror
\begin{minipage}{\linewidth}
\begin{algorithm}[H]
\scriptsize
\SetAlgoLined
 \KwIn{$x$, $m$, $y$:$y_{n-1}y_{n-2}...y_0$}
 \Indp 
\Indm \KwOut{$x^y \mod m$}
\Indp
\Indm
\SetKwProg{myalg}{function}{}{end}
    \BlankLine
    \BlankLine
\Begin{
  $r \gets 1$ \\
  \For{$i \gets n-1$ \textbf{\emph{to}} $0$ } {
    $r\gets \texttt{SQUARE}(r) \mod m$  \\ 
    \If{$y_i = 1$} {
        $r\gets \texttt{MULTIPLY}(r, x) \mod m$
    }
  }
    \KwRet{$r$} 
  } {}

\caption{Square-and-Multiply Modular Exponentiation}
\label{alg:square-multiply}
\end{algorithm}
\end{minipage}
\par
}

{\centering
\removelatexerror
\begin{minipage}{\linewidth}
\begin{algorithm}[H]
\scriptsize
\SetAlgoLined
 \KwIn{$x$, $m$, $y$:$y_{n-1}y_{n-2}...y_0$}
 \Indp 
\Indm \KwOut{$x^y \mod m$}
\Indp
\Indm
\SetKwProg{myalg}{function}{}{end}
    \BlankLine
    \BlankLine
\Begin{
  $r \gets 1$ \\
  \For{$i \gets n-1$ \textbf{\emph{to}} $0$ } {
    $r\gets \texttt{SQUARE}(r) \mod m$  \\ 
    $r'\gets \texttt{MULTIPLY}(r, x) \mod m$ \\
    \If{$y_i = 1$} {
        $r\gets r'$
    }
  }
    \KwRet{$r$} 
  } {}

 \caption{Square-and-Multiply-always Modular Exponentiation}
 \label{alg:square-multiply-always}
\end{algorithm}
\end{minipage}
\par
}

{\centering
\removelatexerror
\begin{minipage}{\linewidth}
\begin{algorithm}[H]
\scriptsize
\SetAlgoLined
 \KwIn{$x$, $m$, $y$:$w_{n-1}w_{n-2}...w_0$}
 \Indp 
\Indm \KwOut{$x^y \mod m$}
\Indp
\Indm
\SetKwProg{myalg}{function}{}{end}
    \BlankLine
    \BlankLine
\Begin{
  $g[0] \gets x \mod m$ \\
  $s \gets \texttt{SQUARE}(g[0]) \mod m$ \\
  \For{$i \gets 1$ \textbf{\emph{to}} $2^{L-1}-1$ } {
    $g[i] \gets \texttt{MULTIPLY}(g[i-1], s) \mod m$ \\
  }

  \BlankLine
  \BlankLine

  $r \gets 1$ \\
  \For{$i \gets n-1$ \textbf{\emph{to}} $0$ } {
      \For{$j \gets 0$ \textbf{\emph{to}} $l_i-1$ } {
        $r\gets \texttt{SQUARE}(r) \mod m$  \\ 
       }
       \If{$w_i \neq 0$} {
          $r\gets \texttt{MULTIPLY}(r, g[(w_i-1)/2]) \mod m$
        }
  }
    \KwRet{$r$} 
  } {}

 \caption{Sliding Window Modular Exponentiation}
 \label{alg:swe}
\end{algorithm}
\end{minipage}
\par
}

{\centering
\removelatexerror
\begin{minipage}{\linewidth}
\begin{algorithm}[H]
\scriptsize
\SetAlgoLined
\Indp
\Indm
\SetKwFunction{proc}{Scatter}
\SetKwFunction{procg}{Gather}
\SetKwProg{myalg}{function}{}{end}
  $s \gets 8$ \\
  $m \gets 32$ \\
  $b \gets 24$ \\
    \BlankLine
    \BlankLine
  \myalg{\proc{$mem, g[i]$}}{

    \For{$j \gets 0$ \textbf{\emph{to}} $b-1$ }{
      \For{$k \gets 0$ \textbf{\emph{to}} $s-1$ } {
          $mem[s*m*j + i*8 + k] \gets g[i][j*s + k]$  \\ 
      }
    }
  }
  
    \BlankLine
    \BlankLine
  \myalg{\procg{$g[i], mem$}}{
  \For{$j \gets 0$ \textbf{\emph{to}} $b-1$ } {
      \For{$k \gets 0$ \textbf{\emph{to}} $s-1$ } {
        $g[i][j*s+k] \gets mem[s*m*j + i*8 + k]$  \\ 
       }
    }
  }
 \caption{Scatter and Gather method}
 \label{alg:scatter-gather}
\end{algorithm}
\end{minipage}
\par
}

{\centering
\removelatexerror
\begin{minipage}{\linewidth}
\begin{algorithm}[H]
\scriptsize
\SetAlgoLined
\Indp
\Indm
\SetKwFunction{procg}{Gather}
\SetKwProg{myalg}{function}{}{end}
  $s \gets 8$ \\
  $m \gets 32$ \\
  $b \gets 24$ \\
    \BlankLine
    \BlankLine
  \myalg{\procg{$g[i], mem$}}{
  \For{$j \gets 0$ \textbf{\emph{to}} $b-1$ } {
    \For{$l \gets 0$ \textbf{\emph{to}} $m-1$}{
      \For{$k \gets 0$ \textbf{\emph{to}} $s-1$ } {
        $v \gets mem[s*m*j + l*8 + k]$  \\         
        $g[i][j*s+k] \gets g[i][j*s+k] | (v \& (m == l)) $  \\ 
       }
    }
   }
  }
 \caption{Masked Window Gather method}
 \label{alg:mask-gather}
\end{algorithm}
\end{minipage}
\par
}

{\centering
\removelatexerror
\begin{minipage}{\linewidth}
\begin{algorithm}[H]
\scriptsize
\SetAlgoLined
 \KwIn{$P$, $N$:$n_{m-1}n_{m-2}...n_0$}
 \Indp 
\Indm \KwOut{$NP$}
\Indp
\Indm
\SetKwProg{myalg}{function}{}{end}
    \BlankLine
    \BlankLine
\Begin{
  $r \gets 0$ \\
  \For{$i \gets m-1$ \textbf{\emph{to}} $0$ } {
    $r\gets \texttt{PointDouble}(r)$  \\ 
    \If{$n_i = 1$} {
        $r\gets \texttt{PointAdd}(r, P)$
    }
  }
    \KwRet{$r$} 
  } {}

\caption{Double-and-Add Scalar Multiplication}
\label{alg:double-and-add}
\end{algorithm}
\end{minipage}
\par
}

{\centering
\removelatexerror
\begin{minipage}{\linewidth}
\begin{algorithm}[H]
\scriptsize
\SetAlgoLined
 \KwIn{$P$, $N$:$n_{m-1}n_{m-2}...n_0$}
 \Indp 
\Indm \KwOut{$NP$}
\Indp
\Indm
\SetKwProg{myalg}{function}{}{end}
    \BlankLine
    \BlankLine
\Begin{
  $r \gets 0$ \\
  \For{$i \gets m-1$ \textbf{\emph{to}} $0$ } {
    $r\gets \texttt{PointDouble}(r)$  \\ 
    $r'\gets \texttt{PointAdd}(r, P)$ \\
    \If{$n_i = 1$} {
        $r \gets r'$
    }
  }
    \KwRet{$r$} 
  } {}

\caption{Double-and-Add-always Scalar Multiplication}
\label{alg:double-and-add-always}
\end{algorithm}
\end{minipage}
\par
}

{\centering
\removelatexerror
\begin{minipage}{\linewidth}
\begin{algorithm}[H]
\scriptsize
\SetAlgoLined
 \KwIn{$P$, $N$}
 \Indp 
\Indm \KwOut{$NP$}
\Indp
\Indm
\SetKwProg{myalg}{function}{}{end}
    \BlankLine
    \BlankLine
\Begin{

  $i \gets 0$ \\
  \While{$N > 0$}{
    \eIf{$N \emph{\text{ mod }} 2 = 1$}{
      \eIf{$N \emph{\text{ mod }} 2^w \geq 2^{w-1}$}{
        $d_i \gets N \text{ mod } 2^w - 2^w$
      }{
        $d_i \gets N \text{ mod } 2^w$
      }
      $N \gets N - d_i$
    }{
      $d_i \gets 0$
    }
    $N \gets N/2$ \\
    $i \gets i + 1$
  }

  \BlankLine
  \BlankLine

  $g[0] \gets P$ \\
  \For{$j \gets 1$ \textbf{\emph{to}} $2^{w-2}-1$ } {
    $g[j] \gets g[j-1] + P$ \\
  }

  \BlankLine
  \BlankLine

  $r \gets 0$ \\
  \For{$j \gets i-1$ \textbf{\emph{to}} $0$ } {
      $r \gets \texttt{PointDouble}(r)$ \\
      \If{$d_j \neq 0$} {
        \eIf{$d_j > 0$}{
          $r \gets \texttt{PointAdd}(r, g[d_j])$
        }{
          $r \gets \texttt{PointAdd}(r, \texttt{Negative}(g[-d_j]))$
        }
      }
  }
    \KwRet{$r$} 
  } {}

 \caption{Sliding Window Scalar Multiplication}
 \label{alg:naf}
\end{algorithm}
\end{minipage}
\par
}

{\centering
\removelatexerror
\begin{minipage}{\linewidth}
\begin{algorithm}[H]
\scriptsize
\SetAlgoLined
 \KwIn{$P$, $N$:$n_{m-1}n_{m-2}...n_0$}
 \Indp 
\Indm \KwOut{$NP$}
\Indp
\Indm
\SetKwProg{myalg}{function}{}{end}
    \BlankLine
    \BlankLine
\Begin{
  $r \gets 0$ \\
  $r' \gets P$ \\
  \For{$i \gets m-1$ \textbf{\emph{to}} $0$ } {
    \eIf{$n_i = 0$} {
        $r'\gets \texttt{PointAdd}(r, r')$ \\
        $r\gets \texttt{PointDouble}(r)$ \\
    }{
        $r\gets \texttt{PointAdd}(r, r')$ \\
        $r'\gets \texttt{PointDouble}(r')$ \\
    }
  }
    \KwRet{$r$} 
  } {}

\caption{Montgomery ladder Scalar Multiplication}
\label{alg:montgomery-ladder}
\end{algorithm}
\end{minipage}
\par
}

{\centering
\removelatexerror
\begin{minipage}{\linewidth}
\begin{algorithm}[H]
\scriptsize
\SetAlgoLined
 \KwIn{$P$, $N$:$n_{m-1}n_{m-2}...n_0$}
 \Indp 
\Indm \KwOut{$NP$}
\Indp
\Indm
\SetKwProg{myalg}{function}{}{end}
    \BlankLine
    \BlankLine
\Begin{
  $r \gets 0$ \\
  $r' \gets P$ \\
  \For{$i \gets m-1$ \textbf{\emph{to}} $0$ } {
    $\texttt{ConstSwap}(r, r', n_i)$ \\
    $r'\gets \texttt{PointAdd}(r, r')$ \\
    $r\gets \texttt{PointDouble}(r)$ \\
    $\texttt{ConstSwap}(r, r', n_i)$
  }
    \KwRet{$r$} 
  } {}

\caption{Branchless montgomery ladder Scalar Multiplication}
\label{alg:branchless-montgomery-ladder}
\end{algorithm}
\end{minipage}
\par
}

{\centering
\removelatexerror
\begin{minipage}{\linewidth}
\begin{algorithm}[H]
\scriptsize
\SetAlgoLined
 \KwIn{$x$, $m$}
 \Indp 
\Indm \KwOut{$x^{-1} \text{ mod } m$}
\Indp
\Indm
\SetKwProg{myalg}{function}{}{end}
    \BlankLine
    \BlankLine
\Begin{
  $v \gets m$ \\
  $u \gets x$ \\
  $p \gets 1$ \\
  $q \gets 0$ \\
  \While{$u \neq 0$}{
    \While{\emph{$u \text{ mod } 2 = 0$}}{
      $u \gets u/2$ \\
      \If{\emph{$p \text{ mod } 2 = 1$}}{
        $p \gets p + m$
      }
      $p \gets p /2$
    }
    \While{\emph{$v \text{ mod } 2 = 0$}}{
      $v \gets v/2$ \\
      \If{\emph{$q \text{ mod } 2 = 1$}}{
        $q \gets q + m$
      }
      $q \gets q /2$
    }
    \eIf{$u \geq v$}{
      $u \gets u - v$ \\
      $p \gets p - q$
    }{
      $v \gets v - u$ \\
      $q \gets q - p$
    }
  }
  $r \gets q \text{ mod } m$ \\
    \KwRet{$r$} 
  } {}
\caption{Binary Extended Euclidean Algorithm}
\label{alg:beea}
\end{algorithm}
\end{minipage}
\par
}

{\centering
\removelatexerror
\begin{minipage}{\linewidth}
\begin{algorithm}[H]
\scriptsize
\SetAlgoLined
 \KwIn{$x$, $m$}
 \Indp 
\Indm \KwOut{$x^{-1} \text{ mod } m$}
\Indp
\Indm
\SetKwProg{myalg}{function}{}{end}
    \BlankLine
    \BlankLine
\Begin{
  $v \gets m$ \\
  $u \gets x$ \\
  $p \gets 1$ \\
  $q \gets 0$ \\
  \While{$u \neq 0$}{
    $tmp_1 \gets v \textbf{ div } u$ \\
    $tmp_2 \gets q - tmp_1 * p$ \\
    $tmp_3 \gets v - tmp_1 * u$ \\
    $q \gets p$ \\
    $p \gets tmp_2$ \\
    $v \gets u$ \\
    $u \gets tmp_3$ \\
  }
  \eIf{$q < 0$}{
    $r \gets q + m$ \\
  }{
    $r \gets q$ \\
  }  
    \KwRet{$r$} 
  } {}
\caption{Extended Euclidean Algorithm}
\label{alg:eea}
\end{algorithm}
\end{minipage}
\par
}

\subsection{CVSS Calculation (version 2.0)}
\label{sec:cvss-cal}
Equation \ref{eq:base} shows the formula of CVSS 2.0 base metrics. The \emph{Base} score
is determined by two sub-scores: \emph{Exploitability} and \emph{Impact}. 

\emph{Exploitability} measures the difficulty to exploit this vulnerability. It is affected
by three factors: (1) Access Vector (\emph{AV}) reflects the location that the attacker needs
to exploit the vulnerability. A farther location is awarded a higher score.
(2) Access Complexity (\emph{AC}) embodies the difficulty of the attack required to exploit the 
vulnerability once the adversary gains access to the target system. Lower complexity is awarded a higher score.
(3) Authentication (\emph{AU}) measures the number of times the adversary
must authenticate to the system in order to exploit the vulnerability. Smaller amount of 
authentications is awarded a higher scores.

\begin{table*}[h!]
\caption{Release history of OpenSSL and GNU Crypto (gray entries contain side-channel patches)}
\label{table:release-history}
\begin{subtable}{0.48\linewidth}
\centering
\resizebox{0.92\linewidth}{!}{
\begin{tabular}{|c|c|c|c|c|c|c|c|c|}
  \hline
\textbf{Date} & \textbf{0.9.6}  & \textbf{0.9.7}  & \textbf{0.9.8}  & \textbf{1.0.0}  & \textbf{1.0.1}  & \textbf{1.0.2}  & \textbf{1.1.0}  & \textbf{1.1.1}  \\
\hline
2000/09/24  & 0.9.6 &   &   &   &   &   &   &   \\
2001/04/05  & 0.9.6a  &   &   &   &   &   &   &   \\
2001/07/09  & \cellcolor{gray}0.9.6b  &   &   &   &   &   &   &   \\
2001/12/21  & 0.9.6c  &   &   &   &   &   &   &   \\
2002/05/08  & 0.9.6d  &   &   &   &   &   &   &   \\
2002/07/30  & 0.9.6e  &   &   &   &   &   &   &   \\
2002/08/08  & 0.9.6f  &   &   &   &   &   &   &   \\
2002/08/09  & 0.9.6g  &   &   &   &   &   &   &   \\
2002/12/05  & 0.9.6h  &   &   &   &   &   &   &   \\
\cline{3-3}
2002/12/31  &   & 0.9.7 &   &   &   &   &   &   \\
2003/02/19  & \cellcolor{gray}0.9.6i  & \cellcolor{gray}0.9.7a  &   &   &   &   &   &   \\
2003/04/10  & \cellcolor{gray}0.9.6j  & \cellcolor{gray}0.9.7b  &   &   &   &   &   &   \\
2003/09/30  & 0.9.6k  & 0.9.7c  &   &   &   &   &   &   \\
2003/11/04  & 0.9.6l  &   &   &   &   &   &   &   \\
2004/03/17  &   & 0.9.7d  &   &   &   &   &   &   \\
2004/05/17  & 0.9.6m  &   &   &   &   &   &   &   \\
\cline{2-2}
2004/10/25  &   & 0.9.7e  &   &   &   &   &   &   \\
2005/03/22  &   & 0.9.7f  &   &   &   &   &   &   \\
2005/04/11  &   & 0.9.7g  &   &   &   &   &   &   \\
\cline{4-4}
2005/07/05  &   &   & \cellcolor{gray}0.9.8 &   &   &   &   &   \\
2005/10/11  &   & \cellcolor{gray}0.9.7h  & 0.9.8a  &   &   &   &   &   \\
2005/10/14  &   & 0.9.7i  &   &   &   &   &   &   \\
2006/05/04  &   & 0.9.7j  & 0.9.8b  &   &   &   &   &   \\
2006/09/05  &   & 0.9.7k  & 0.9.8c  &   &   &   &   &   \\
2006/09/28  &   & 0.9.7l  & 0.9.8d  &   &   &   &   &   \\
\cline{3-3}
2007/02/23  &   &   & 0.9.8e  &   &   &   &   &   \\
2007/10/11  &   &   & \cellcolor{gray}0.9.8f  &   &   &   &   &   \\
2007/10/19  &   &   & 0.9.8g  &   &   &   &   &   \\
2008/05/28  &   &   & 0.9.8h  &   &   &   &   &   \\
2008/09/15  &   &   & 0.9.8i  &   &   &   &   &   \\
2009/01/07  &   &   & 0.9.8j  &   &   &   &   &   \\
2009/05/25  &   &   & 0.9.8k  &   &   &   &   &   \\
2009/11/05  &   &   & 0.9.8l  &   &   &   &   &   \\
2010/02/25  &   &   & 0.9.8m  &   &   &   &   &   \\
2010/03/24  &   &   & 0.9.8n  &   &   &   &   &   \\
\cline{5-5}
2010/03/29  &   &   &   & 1.0.0 &   &   &   &   \\
2010/06/01  &   &   & 0.9.8o  & 1.0.0a  &   &   &   &   \\
2010/11/16  &   &   & 0.9.8p  & 1.0.0b  &   &   &   &   \\
2010/12/02  &   &   & 0.9.8q  & 1.0.0c  &   &   &   &   \\
2011/02/08  &   &   & 0.9.8r  & 1.0.0d  &   &   &   &   \\
2011/09/06  &   &   &   & \cellcolor{gray}1.0.0e  &   &   &   &   \\
2012/01/04  &   &   & \cellcolor{gray}0.9.8s  & \cellcolor{gray}1.0.0f  &   &   &   &   \\
2012/01/18  &   &   & 0.9.8t  & 1.0.0g  &   &   &   &   \\
2012/03/12  &   &   & \cellcolor{gray}0.9.8u  & \cellcolor{gray}1.0.0h  &   &   &   &   \\
\cline{6-6}
2012/03/14  &   &   &   &   & \cellcolor{gray}1.0.1 &   &   &   \\
2012/04/19  &   &   & 0.9.8v  & 1.0.0i  & 1.0.1a  &   &   &   \\
2012/04/23  &   &   & 0.9.8w  &   &   &   &   &   \\
2012/04/26  &   &   &   &   & 1.0.1b  &   &   &   \\
2012/05/10  &   &   & 0.9.8x  & 1.0.0j  & 1.0.1c  &   &   &   \\
2013/02/05  &   &   & \cellcolor{gray}0.9.8y  & \cellcolor{gray}1.0.0k  & \cellcolor{gray}1.0.1d  &   &   &   \\
2013/02/11  &   &   &   &   & 1.0.1e  &   &   &   \\
2014/01/06  &   &   &   & 1.0.0l  & 1.0.1f  &   &   &   \\
2014/04/07  &   &   &   &   & \cellcolor{gray}1.0.1g  &   &   &   \\
2014/06/05  &   &   & \cellcolor{gray}0.9.8za & \cellcolor{gray}1.0.0m  & 1.0.1h  &   &   &   \\
2014/08/06  &   &   & 0.9.8zb & 1.0.0n  & 1.0.1i  &   &   &   \\
2014/10/15  &   &   & \cellcolor{gray}0.9.8zc & \cellcolor{gray}1.0.0o  & \cellcolor{gray}1.0.1j  &   &   &   \\
2015/01/08  &   &   & 0.9.8zd & 1.0.0p  & 1.0.1k  &   &   &   \\
2015/01/15  &   &   & 0.9.8ze & 1.0.0q  & 1.0.1l  &   &   &   \\
\cline{7-7}
2015/01/22  &   &   &   &   &   & 1.0.2 &   &   \\
2015/03/19  &   &   & 0.9.8zf & 1.0.0r  & 1.0.1m  & 1.0.2a  &   &   \\
2015/06/11  &   &   & 0.9.8zg & 1.0.0s  & 1.0.1n  & 1.0.2b  &   &   \\
\cline{5-5}
2015/06/12  &   &   &   &   & 1.0.1o  & 1.0.2c  &   &   \\
2015/07/09  &   &   &   &   & 1.0.1p  & 1.0.2d  &   &   \\
2015/12/03  &   &   & 0.9.8zh &   & 1.0.1q  & 1.0.2e  &   &   \\
\cline{4-4}
2016/01/28  &   &   &   &   & \cellcolor{gray}1.0.1r  & \cellcolor{gray}1.0.2f  &   &   \\
2016/03/01  &   &   &   &   & \cellcolor{gray}1.0.1s  & \cellcolor{gray}1.0.2g  &   &   \\
2016/03/03  &   &   &   &   & \cellcolor{gray}1.0.1t  & \cellcolor{gray}1.0.2h  &   &   \\
2016/08/25  &   &   &   &   &   &   & 1.1.0 &   \\
\cline{8-8}
2016/09/22  &   &   &   &   & \cellcolor{gray}1.0.1u  & \cellcolor{gray}1.0.2i  & 1.1.0a  &   \\
\cline{6-6}
2016/09/26  &   &   &   &   &   & 1.0.2j  & 1.1.0b  &   \\
2016/11/10  &   &   &   &   &   &   & 1.1.0c  &   \\
2017/01/26  &   &   &   &   &   & 1.0.2k  & 1.1.0d  &   \\
2017/02/16  &   &   &   &   &   &   & 1.1.0e  &   \\
2017/05/25  &   &   &   &   &   & 1.0.2l  & 1.1.0f  &   \\
2017/11/02  &   &   &   &   &   & 1.0.2m  & 1.1.0g  &   \\
2017/12/07  &   &   &   &   &   & 1.0.2n  &   &   \\
2018/03/27  &   &   &   &   &   & 1.0.2o  & 1.1.0h  &   \\
2018/08/14  &   &   &   &   &   & \cellcolor{gray}1.0.2p  & \cellcolor{gray}1.1.0i  &   \\
\cline{9-9}
2018/09/11  &   &   &   &   &   &   &   & \cellcolor{gray}1.1.1 \\
2018/11/20  &   &   &   &   &   & \cellcolor{gray}1.0.2q  & \cellcolor{gray}1.1.0j  & \cellcolor{gray}1.1.1a  \\
2019/02/26  &   &   &   &   &   & \cellcolor{gray}1.0.2r  &   & \cellcolor{gray}1.1.1b  \\
\hline
  \end{tabular}
}
\caption{OpenSSL}
\label{table:openssl-release}
\end{subtable}%
\begin{subtable}{0.48\linewidth}
\centering
\resizebox{0.92\linewidth}{!}{
\begin{tabular}{|c|c|c|c|c|c|c|c|c|}
  \hline
\textbf{Date} & \textbf{P1.4.x}  & \textbf{P2.0.x}  & \textbf{P2.1.x}  & \textbf{P2.2.x}  & \textbf{L1.5.x}  & \textbf{L1.6.x}  & \textbf{L1.7.x}  & \textbf{L1.8.x}  \\
\hline
2011/06/29  &   &   &   &   & \cellcolor{gray}1.5.0 &   &   &   \\
2012/12/20  & 1.4.13  &   &   &   &   &   &   &   \\
2013/03/18  &   &   &   &   & 1.5.1 &   &   &   \\
2013/04/18  &   &   &   &   & 1.5.2 &   &   &   \\
2013/05/10  &   & 2.0.20  &   &   &   &   &   &   \\
2013/07/25  & \cellcolor{gray}1.4.14  &   &   &   & \cellcolor{gray}1.5.3 &   &   &   \\
2013/08/19  &   & 2.0.21  &   &   &   &   &   &   \\
2013/10/04  & 1.4.15  & 2.0.22  &   &   &   &   &   &   \\
\cline{7-7}
2013/12/16  &   &   &   &   &   & \cellcolor{gray}1.6.0 &   &   \\
2013/12/18  & \cellcolor{gray}1.4.16  &   &   &   &   &   &   &   \\
2014/01/29  &   &   &   &   &   & 1.6.1 &   &   \\
2014/06/03  &   & 2.0.23  &   &   &   &   &   &   \\
2014/06/23  & 1.4.17  &   &   &   &   &   &   &   \\
2014/06/24  &   & 2.0.24  &   &   &   &   &   &   \\
2014/06/30  & 1.4.18  & 2.0.25  &   &   &   &   &   &   \\
2014/08/07  &   &   &   &   & \cellcolor{gray}1.5.4 &   &   &   \\
2014/08/12  &   & 2.0.26  &   &   &   &   &   &   \\
2014/08/21  &   &   &   &   &   & 1.6.2 &   &   \\
2014/11/06  &   &   & 2.1.0 &   &   &   &   &   \\
2014/12/16  &   &   & 2.1.1 &   &   &   &   &   \\
2015/02/11  &   &   & 2.1.2 &   &   &   &   &   \\
2015/02/18  &   & 2.0.27  &   &   &   &   &   &   \\
2015/02/27  & \cellcolor{gray}1.4.19  &   &   &   &   & \cellcolor{gray}1.6.3 &   &   \\
2015/04/11  &   &   & 2.1.3 &   &   &   &   &   \\
2015/05/12  &   &   & 2.1.4 &   &   &   &   &   \\
2015/06/02  &   & 2.0.28  &   &   &   &   &   &   \\
2015/06/11  &   &   & 2.1.5 &   &   &   &   &   \\
2015/07/01  &   &   & 2.1.6 &   &   &   &   &   \\
2015/08/11  &   &   & 2.1.7 &   &   &   &   &   \\
2015/09/08  &   & 2.0.29  &   &   &   & 1.6.4 &   &   \\
2015/09/10  &   &   & 2.1.8 &   &   &   &   &   \\
2015/10/09  &   &   & 2.1.9 &   &   &   &   &   \\
2015/12/04  &   &   & 2.1.10  &   &   &   &   &   \\
2015/12/20  & 1.4.20  &   &   &   &   &   &   &   \\
2016/01/26  &   &   & 2.1.11  &   &   &   &   &   \\
2016/02/09  &   &   &   &   &   & \cellcolor{gray}1.6.5 &   &   \\
2016/02/18  &   &   &   &   & \cellcolor{gray}1.5.5 &   &   &   \\
2016/03/31  &   & 2.0.30  &   &   &   &   &   &   \\
\cline{8-8}
2016/04/15  &   &   &   &   &   &   & \cellcolor{gray}1.7.0 &   \\
2016/05/04  &   &   & 2.1.12  &   &   &   &   &   \\
2016/06/15  &   &   &   &   &   &   & 1.7.1 &   \\
2016/06/16  &   &   & 2.1.13  &   &   &   &   &   \\
2016/07/14  &   &   & 2.1.14  &   &   &   & 1.7.2 &   \\
2016/08/17  & 1.4.21  &   &   &   & 1.5.6 & 1.6.6 & 1.7.3 &   \\
\cline{6-7}
2016/08/18  &   &   & 2.1.15  &   &   &   &   &   \\
2016/11/18  &   &   & 2.1.16  &   &   &   &   &   \\
2016/12/09  &   &   &   &   &   &   & 1.7.4 &   \\
2016/12/15  &   &   &   &   &   &   & 1.7.5 &   \\
2016/12/20  &   &   & 2.1.17  &   &   &   &   &   \\
2017/01/18  &   &   &   &   &   &   & 1.7.6 &   \\
2017/01/23  &   &   & 2.1.18  &   &   &   &   &   \\
2017/03/01  &   &   & 2.1.19  &   &   &   &   &   \\
2017/04/03  &   &   & 2.1.20  &   &   &   &   &   \\
2017/05/15  &   &   & 2.1.21  &   &   &   &   &   \\
2017/06/02  &   &   &   &   &   &   & 1.7.7 &   \\
2017/06/29  &   &   &   &   &   &   & \cellcolor{gray}1.7.8 &   \\
\cline{9-9}
2017/07/18  &   &   &   &   &   &   &   & \cellcolor{gray}1.8.0 \\
2017/07/19  & \cellcolor{gray}1.4.22  &   &   &   &   &   &   &   \\
2017/07/28  &   &   & 2.1.22  &   &   &   &   &   \\
2017/08/09  &   &   & 2.1.23  &   &   &   &   &   \\
\cline{4-4}
2017/08/27  &   &   &   &   &   &   & \cellcolor{gray}1.7.9 & \cellcolor{gray}1.8.1 \\
2017/08/28  &   &   &   & 2.2.0 &   &   &   &   \\
2017/09/19  &   &   &   & 2.2.1 &   &   &   &   \\
2017/11/07  &   &   &   & 2.2.2 &   &   &   &   \\
2017/11/20  &   &   &   & 2.2.3 &   &   &   &   \\
2017/12/13  &   &   &   &   &   &   &   & 1.8.2 \\
2017/12/20  &   &   &   & 2.2.4 &   &   &   &   \\
2017/12/29  &   & 2.0.31  &   &   &   &   &   &   \\
\cline{3-3}
2018/02/22  &   &   &   & 2.2.5 &   &   &   &   \\
2018/04/09  &   &   &   & 2.2.6 &   &   &   &   \\
2018/05/02  &   &   &   & 2.2.7 &   &   &   &   \\
2018/06/08  &   &   &   & 2.2.8 &   &   &   &   \\
2018/06/11  & 1.4.23  &   &   &   &   &   &   &   \\
2018/06/13  &   &   &   &   &   &   & \cellcolor{gray}1.7.10  & \cellcolor{gray}1.8.3 \\
2018/07/12  &   &   &   & 2.2.9 &   &   &   &   \\
2018/08/30  &   &   &   & 2.2.10  &   &   &   &   \\
2018/10/26  &   &   &   &   &   &   &   & 1.8.4 \\
2018/11/06  &   &   &   & 2.2.11  &   &   &   &   \\
2018/12/14  &   &   &   & 2.2.12  &   &   &   &   \\
2019/02/12  &   &   &   & 2.2.13  &   &   &   &   \\
\hline
  \end{tabular}
}
  \caption{GNU Crypto}
  \label{table:gnupg-release}
  \end{subtable}
\vspace{-15pt}
\end{table*}

\emph{Impact} measures how much damage the vulnerability can incur to the target system. It
is evaluated in terms of three security properties: (1) Confidentiality (\emph{CImpact}) 
refers to the amount of information leaked to the adversary. 
(2) Integrity (\emph{IImpact}) refers to the amount of data that the adversary can tamper with. 
(3) Availability (\emph{AImpact}) measures the loss of access to the system information, 
resources and services. 
The possible values of the three metrics can be none, partial breach, and complete breach.

\begin{align}
\label{eq:base}
\scriptsize
\begin{split}
& Base = 
\begin{cases}
    0, \qquad \text{if }Impact = 0\\
    (0.6*Impact+0.4*Exploitability-1.5)*1.176, \text{otherwise}
\end{cases} \\
& Exploitability = 20*AV*AC*AU \\
& Impact = 10.41*[1-(1-CImpact)*(1-IImpact)*(1-AImpact)] \\
& AV = 0.395\text{ (local) }/ 0.646\text{ (adjacent network) }/ 1\text{ (network) }\\
& AC = 0.35\text{ (high) }/ 0.61\text{ (medium) }/ 0.71\text{ (low) }\\
& AU = 0.45\text{ (multiple) }/ 0.56\text{ (single) }/ 0.704\text{ (no) }\\
& CImpact = 0\text{ (none) }/ 0.275\text{ (partial) }/ 0.66\text{ (complete) }\\
& IImpact = 0\text{ (none) }/ 0.275\text{ (partial) }/ 0.66\text{ (complete) }\\
& AImpact =  0\text{ (none) }/ 0.275\text{ (partial) }/ 0.66\text{ (complete) }\\
\end{split}
\end{align}

\subsection{Release history of cryptographic libraries}
\label{sec:release-lib}

Table \ref{table:release-history} shows the release history of OpenSSL and GNU Crypto libraries for
the past years. We highlight the released versions containing side-channel patches in gray.

For OpenSSL, we observe that it keeps maintaining about three live branches concurrently throughout
the history. The patches of most vulnerabilities were applied to all live branches at the same time. 
Thus, OpenSSL has good cross-branch patch consistency for side-channel vulnerabilities.

The case of GNU Crypto is more complicated. We ignore GnuTLS as it has too many branches and versions. 
Libgcrypt was previously a module inside GnuPG for cryptographic primitives, but later detached itself 
to become an independent library. As a result, some GnuPG branches (1.4) continued to keep
this module, while others (2.0, 2.1) did not. Also, some implementations of the same
operations in GnuPG and Libgcrypt differed significantly. Thus, side-channel patches
across libraries and branches appeared fairly inconsistent.

\subsection*{Disclaimer}

The views expressed herein are solely the views of the author(s) and are not necessarily the views of
Two Sigma Investments, LP or any of its affiliates. They are not intended to provide, and should not be
relied upon for, investment advice.
\end{document}